\documentclass[letterpaper,twocolumn,10pt]{article}
\usepackage{graphicx}
\usepackage{balance}  % for  \balance command ON LAST PAGE  (only there!)

\usepackage{epsfig,endnotes}
\usepackage{cite}
\usepackage{amsmath,amssymb,amsfonts}
\usepackage{algorithmic}
\usepackage{graphicx}
\usepackage{textcomp}
\usepackage{amsfonts,amsmath,amssymb}
\usepackage{framed}
\usepackage{array}
\usepackage{epsfig}
\usepackage{tikz}
\usepackage{enumitem}
\usepackage{url}
\usepackage{graphicx}
\usepackage{pgfplotstable}
\usepackage{pgfplots}
\usepackage{microtype}
\usepackage{tabularx}
\usepackage{color}
\usepackage{times}
\usepackage[small,compact]{titlesec}

\usepackage{booktabs}
\usepackage{threeparttable}
\usepackage{wasysym}

\newenvironment{denseitemize}{
\begin{itemize}[topsep=2pt, partopsep=0pt, leftmargin=1.5em]
  \setlength{\itemsep}{4pt} 
  \setlength{\parskip}{0pt}
  \setlength{\parsep}{0pt}
}{\end{itemize}}

\newenvironment{denseenumerate}{
\begin{enumerate}[topsep=2pt, partopsep=0pt, leftmargin=1.5em]
  \setlength{\itemsep}{4pt}
  \setlength{\parskip}{0pt}
  \setlength{\parsep}{0pt}
}{\end{enumerate}}

\newcommand{\ignore}[1]{}
\newtheorem{theorem}{Theorem}
\newtheorem{definition}[theorem]{Definition}

\def\name/{ObliDB}

\newcommand{\rev}[1]{#1}

\def\theTitle{ObliDB: Oblivious Query Processing for Secure Databases}

\pgfplotsset{width=6cm,compat=1.9}

\begin{document}

% ****************** TITLE ****************************************

\title{\textbf{\theTitle}}

% possible, but not really needed or used for PVLDB:
%\subtitle{[Extended Abstract]
%\titlenote{A full version of this paper is available as\textit{Author's Guide to Preparing ACM SIG Proceedings Using \LaTeX$2_\epsilon$\ and BibTeX} at \texttt{www.acm.org/eaddress.htm}}}

\author{
{\rm Saba Eskandarian}\\
Stanford University
\and
{\rm Matei Zaharia}\\
Stanford University
% copy the following lines to add more authors
% \and
% {\rm Name}\\
%Name Institution
} % 
\date{}
% Just remember to make sure that the TOTAL number of authors
% is the number that will appear on the first page PLUS the
% number that will appear in the \additionalauthors section.

\maketitle

\begin{abstract}
Hardware enclaves such as Intel SGX are a promising technology for improving the security of databases outsourced to the cloud.
These enclaves provide an execution environment isolated from the hypervisor/OS, and encrypt data in RAM.
However, for applications that use large amounts of memory, including most databases, enclaves do not protect against \emph{access pattern} leaks, which let attackers gain a large amount of information about the data.
Moreover, the na\"ive way to address this issue, using Oblivious RAM (ORAM) primitives from the security literature, adds substantial overhead.

A number of recent works explore trusted hardware enclaves as a path toward secure, access-pattern oblivious outsourcing of data storage and analysis. While these works efficiently solve specific subproblems (e.g. building secure indexes or running analytics queries that always scan entire tables), no prior work has supported oblivious query processing for \rev{general query workloads on a DBMS engine with multiple access methods}. 
\rev{Moreover, applying these techniques individually does not guarantee that an end-to-end workload, such as a complex SQL query over multiple tables, will be oblivious.} 
\rev{In this paper, we introduce \name/, an oblivious database engine design that is the first system to provide obliviousness for general database read workloads over multiple access methods}.

%both transactional and analytic workloads.

\name/ introduces a diverse array of new oblivious physical operators to accelerate oblivious SQL queries, giving speedups of up to an order of magnitude over na\"ive ORAM.
It supports a broad range of queries, including aggregation, joins, insertions, deletions and point queries. 
%Crucially, \name/ ensures that queries using multiple of its operators also remain oblivious. 
We implement \name/ and show that, on analytics workloads, \name/ ranges from 1.1--19$\times$ faster than Opaque, a previous oblivious, enclave-based system designed \emph{only} for analytics, and comes within 2.6$\times$ of Spark SQL, which provides no security guarantees.
In addition, \name/ supports point queries with 3--10ms latency, which is comparable to index-only trusted hardware systems, and runs over 7$\times$ faster than HIRB, a previous encryption-based oblivious index system that supports point queries.
\end{abstract}
\section{Introduction}

\ignore{
\begin{table}\label{compTable}
\center
\tabcolsep=0.11cm
\footnotesize
\newcommand*\rot[1]{\hbox to1em{\hss\rotatebox[origin=br]{-45}{#1}}}
\newcommand*\f[1]{\ifcase#1 -\or$\times$\or\checkmark\fi}
\begin{threeparttable}
\begin{tabular}{lcccccccc}
\toprule
Properties & Oblivious Systems \\
\midrule
&\rot{\name/}            
&\rot{Oblix~\cite{oblix}}           
&\rot{POSUP~\cite{POSUP}}            
&\rot{Cipherbase~\cite{cipherbase}}  
&\rot{Opaque~\cite{ZDB+17}}          
&\rot{Obladi~\cite{obladi}}           
&\rot{StealthDB~\cite{stealthDB}} 
%&\rot{EnclaveDB~\cite{enclavedb}}    
\\
\midrule
Oblivious Indexes              &\f{2}&\f{2}&\f{2}&\f{1}&\f{1}&\f{1}&\f{1}\\%&\f{1} \\ 
Oblivious Analytics            &\f{2}&\f{1}&\f{1}&\f{2}&\f{2}&\f{2}&\f{2}\\%&\f{1} \\
%Query Processing               &\f{2}&\f{1}&\f{1}&\f{2}&\f{2}&\f{2}&\f{2}&\f{2} \\
%General workloads              &\f{2}&\f{1}&\f{1}&\f{1}&\f{2}&\f{2}&\f{2}&\f{2} \\
%Handles requests continuously  &\f{1}&\f{1}&\f{1}&\f{1}&\f{1}&\f{1}&\f{1}&\f{1} \\
\bottomrule
\end{tabular}
\end{threeparttable}
\caption{\name/ is the first system to provide oblivious query processing for both transactional and analytic workloads.}
\end{table}
}

Many organizations outsource their databases to the public cloud to take advantage of its cost efficiency, high availability, and convenience.
Due to the sensitivity of this data, both users and cloud providers would like strong privacy and security guarantees, ideally protecting against both external attackers and insiders that breach the cloud provider's security~\cite{yahoo-hack,nsa-spy,atos-hack}.
To address this problem, researchers have proposed approaches including property preserving encryption~\cite{PRZB12,mylar,FVY+17}, trusted hardware~\cite{trusteddb,cipherbase,ZDB+17}, and algorithms to run specific computations securely~\cite{WZPM16,NWI+13,WYG+17}, giving various tradeoffs between security, generality, and performance.

One of the most promising practical approaches to increase security is the hardware enclave~\cite{CD16,CLD16}.
These enclaves provide an environment where a remotely verifiable piece of code can run without interference from the hypervisor and OS, accessing a small amount of private enclave memory and making upcalls to the operating system for I/O.
Increasing availability of hardware enclaves has further spurred interest in strong cloud security guarantees~\cite{CD16,CLD16}.
Enclaves are already available on many recent CPUs~\cite{CD16,arm-trustzone} and will soon be offered on Microsoft and Google's public clouds~\cite{azure-confidential,asylo}, making them a powerful technology to investigate for secure database hosting~\cite{enclavedb}.

Unfortunately, although enclaves are powerful, they leave open one key threat: \emph{access pattern} attacks.
Applications that use an enclave to manage large amounts of data must still access data through the OS (e.g., to read new memory pages into the enclave or access the disk), so an attacker that controls the OS can see the pattern of addresses being accessed. This leaks a great deal of information, allowing attackers to learn details of both the data itself and users' queries on the data~\cite{IKK12,XCP15,mr-leakage,GMN+16}. The special case of encrypted databases has a long history of surprising leakage at the hands of access pattern attacks~\cite{GMN+16,KKN+16,AAG17,CGPR15,GAB+17,LZWT14,PW16,ZKP16}.

In response to this threat, research has begun to press toward the goal of general-purpose \emph{oblivious} (access pattern-hiding) SQL databases using hardware enclaves. The generic approach to establishing obliviousness uses Oblivious RAM (ORAM)~\cite{GO96,SDS+13}, which guarantees that any two sets of access patterns are indistinguishable from each other, so long as they are \emph{of the same length}. Unfortunately, conventional query processing algorithms vary both the addresses and total number of memory accesses depending on data and queries, rendering generic use of ORAM alone insufficient. 
POSUP~\cite{POSUP} and Oblix~\cite{oblix} explore oblivious indexes over encrypted data using specialized ORAM constructions as building blocks, but do not support general queries. Moreover, oblivious indexes alone do not fully solve the security problem: thus, an attacker can see \emph{how many accesses to an index} occurred during a query operator. 

On the other hand, Cipherbase~\cite{cipherbase} and Opaque~\cite{ZDB+17} propose schemes that hide access patterns, but they are limited to workloads that \emph{scan entire tables}. For example, Opaque relies on oblivious sorts over the entire dataset. These systems are not efficient for more general workloads that may also include point queries. Attempts to support general workloads, such as Obladi~\cite{obladi} and StealthDB~\cite{stealthDB}, also lack key features -- Obladi does not support indexes and requires operations to be processed in batches, and StealthDB does not provide integrity or hide access patterns to indexes.
Thus, prior solutions do not provide algorithms for a \rev{general-purpose DBMS that combines queries of varying selectivity}, the typical use case for outsourced databases (e.g. MySQL, Postgres, etc). %Table~\ref{compTable} summarizes this comparison of prior work.

\noindent\textbf{Our contributions}. This paper introduces \name/, \rev{the first engine to provide efficient, oblivious read queries for relational workloads over multiple access methods}.
The key contribution is a set of oblivious query processing algorithms that work efficiently over both entire datasets and small subsets of data, closing the gap between prior work and general-purpose databases. Often the direct port of a standard operator into an oblivious version is not only slow but also inherently leaky. Our algorithms take advantage of knowledge about query selectivity to maintain obliviousness while outperforming na\"ive oblivious versions of standard techniques. For example, we offer four oblivious \textsf{SELECT} algorithms that vary their interaction with trusted/untrusted memory to achieve obliviousness while optimizing performance for different settings. Our algorithms only leak the structure of queries (hiding parameters) and the size of the output data, the same as Opaque's oblivious mode~\cite{ZDB+17}\footnote{
    We also support padding intermediate and final results of complex queries to a fixed size, similar to Opaque's pad mode~\cite{ZDB+17}, if desired.
}.

\name/'s performance improves over prior systems by supporting multiple storage methods and including a query planner that obliviously chooses the best option among several algorithms to satisfy a given query.
Unlike prior work, \name/ provides two storage methods for its tables: a ``flat'' one, where the table is encrypted as a contiguous file and always scanned (as in Opaque and Cipherbase), and an \emph{oblivious B+ tree} built over ORAM but modified to prevent leakage and performance penalties involved in a direct composition of B+ trees and ORAM. In particular, we hide the path taken in an index to retrieve records as well as the changes made to index data structures on insertions and deletions. 
Each table can be stored using one or both methods, similarly to how administrators can decide to create indexes in traditional databases. For instance, if a table is stored using both methods, \name/ can use the index for point queries and the flat table for full-table aggregation queries.

Choosing between several algorithms to satisfy a query opens the possibility of leaking information about queries or data through algorithm choice. \name/'s query planner mitigates this risk by basing optimization decisions on information already available to the attacker, such as table and query result sizes.% (except when running in padding mode). 

These features let \name/ support a wide range of queries efficiently and securely.
\name/ supports selections, aggregations and joins, as well as efficient point and small range lookups, insertions, deletions, and updates. \rev{Since \name/'s focus is on obliviously processing read queries, the engine does not provide full support for transactions, but techniques for concurrency and logging~\cite{obladi} can be added on top of \name/'s algorithms and storage methods.}

We implement \name/ over Intel SGX~\cite{CD16} and evaluate it on diverse applications and find that it outperforms previous oblivious systems and achieves practical performance compared to systems with no security guarantees.
For analytics, we compare \name/ to Opaque~\cite{ZDB+17} on the Big Data Benchmark~\cite{BDB} and find that it is competitive with Opaque on most queries, but can outperform Opaque by 19$\times$ on queries that can leverage indexes.
\name/ also comes within 2.6$\times$ of Spark SQL~\cite{SparkSQL}, which provides no security guarantees.
For \rev{point queries}, we compare to an open-source encrypted index and find that \name/ outperforms the HIRB + vORAM of Roche et al.~\cite{RAC16} by over 7$\times$.
Moreover, point insertions, deletions and selects \rev{using \name/'s indexes} on a 1M row dataset take 3.6--9.4ms, which is acceptable for many applications and comparable to the other enclave-based indexes Oblix~\cite{oblix} and POSUP~\cite{POSUP} that do not support the more general queries handled by \name/.
%We also compare \name/ to a baseline where existing database algorithms are generically modified to run over ORAM, and show that \name/ can outperform na\"ive use of ORAM by orders of magnitude.
Finally, we show that the choice of physical operators in \name/ enables meaningful query optimization, yielding speedups of up to $11\times$. \name/ is open source at {\small\url{https://github.com/SabaEskandarian/ObliDB}}.

To summarize, our contributions are:
\begin{denseitemize}
\item Oblivious query processing algorithms optimized to run over both indexed and unstructured data, suitable for general purpose SQL databases.
\item The design of \name/, an enclave-based oblivious database engine that \rev{efficiently runs general relational read workloads over multiple access methods}.%, using a combination of flat tables and oblivious B+ trees.
\item A lightweight query planner to choose between operator implementations offered by \name/. 
\item An implementation and evaluation of \name/ using Intel SGX.
\end{denseitemize}

\ignore{
\noindent\textbf{Hardware Enclave Security}. Our implementation of \name/ is built on Intel's SGX, which has recently been shown to be vulnerable to several side-channel attacks~\cite{sgxpectre,foreshadow} based on speculative execution. Although it does not appear that the current version of SGX satisfies the security properties desired of a hardware enclave, Intel is updating SGX and it is likely future versions of SGX can be made secure enough to provide meaningful security guarantees. We stress, however, that other enclave architectures are also available, e.g. the RISC-V based Sanctum~\cite{CLD16}, and that the design of \name/ can be implemented over any trusted hardware platform. 
}

\section{Background and Security Goals}\label{model}
This section gives background on hardware enclaves, describes our threat model, and states our desired security properties. The fundamental goal of \name/ is to protect both user data and query parameters from a malicious attacker with full power to manipulate components of the system lying outside a trusted hardware enclave. This includes protection against both direct observation/modification of data and indirect observation of access pattern leakage.

\subsection{Background}
A \emph{hardware enclave} provides developers with the abstraction of a secure portion of the processor that can verifiably run a trusted code base (TCB) and protect its limited memory from a malicious or compromised OS \cite{CD16, SGXRef}. Developers get a small memory region hidden from the OS and cleared when execution enters or exits an enclave. In this memory, the trusted code can keep secrets from an untrusted OS that otherwise controls the machine. 
The hardware handles the process of entering and exiting an enclave and hiding the activity of the enclave while non-enclave code runs. Enclave code may require access to OS resources such as networking and I/O, so developers specify an interface between the enclave and OS. 

An enclave proves that it runs an untampered version of the desired code through an \textit{attestation} mechanism. Attestation involves an enclave providing a signed hash of its initial state (including the running code), which a client compares with the expected value and rejects if there is any evidence of a corrupted program. 

\subsection{Threat Model}
We leverage a trusted hardware enclave to protect against an attacker with full control of the operating system (OS). We assume that our attacker has the power to examine and modify untrusted memory, network communication, and communication between the processor and enclave. Moreover, it can observe access patterns to untrusted memory and maliciously interrupt the execution of an enclave. We note that an OS-level attacker can always launch an indefinite denial of service attack against an enclave, but such an attack does not compromise privacy. We also allow our attacker to use arbitrary auxiliary information about the nature of data stored. For example, if a database is storing patient data, this includes the incidence of various diseases in the general population.

We assume the security of the trusted hardware platform in that the enclave hides contents of its protected memory pages and CPU registers from an attacker with control of the OS and the attacker cannot subvert the remote attestation process by which the enclave proves its authenticity. Power analysis and timing side channels are out of the scope.
Furthermore, we assume a secure channel exists through which a user can send messages to the enclave: for example, a client can establish such a connection to the enclave through TLS.

We implement our techniques on Intel's SGX~\cite{CD16} due to its popularity and widespread availability. Although several side-channel attacks based on abusing page faults, branching history, or speculative execution have been demonstrated against SGX's protected memory~\cite{BMD+17,LSG+16,XCP15,foreshadow,sgxpectre,WKPK16}, mitigations exist to handle some of these attacks~\cite{SCNS16,RLT15,SLK+17,SLKP17}, and other hardware enclave designs avoid the pitfalls that leave SGX vulnerable~\cite{CLD16, LHM+15, MLS+13}. In particular, the RISC-V based Sanctum~\cite{CLD16} provides a developer abstraction similar to SGX with minimal performance overhead.

\noindent\textbf{Limited Oblivious Memory}.
 We assume a limited amount of oblivious memory is available to the enclave and protected from access pattern leaks (as in Opaque~\cite{ZDB+17}, to which we compare). That is, when the enclave makes a memory access inside this region, the operating system cannot determine which part was accessed. We note that SGX does not provide this kind of obliviousness. However, other similar enclave designs such as Sanctum or RISC-V's Keystone do provide it with little additional overhead, and the principles of the ObliDB system can run just as well on any other enclave architecture. Moreover, many of our oblivious operators, including the query planner, all \textsf{SELECT} algorithms except the ``Small'' algorithm, \rev{and one of our \textsf{JOIN} algorithms,} maintain obliviousness even with an enclave completely vulnerable to these attacks, i.e. with 0MB of oblivious memory. The quantity of oblivious memory can be set as small as a few megabytes. It primarily serves to store the root position map for our ORAM implementation, and is also used to improve performance for the aggregation, grouped aggregation, and join operators (Section~\ref{oblivOps}) and hide accesses to code pages. The amount of oblivious memory can be reduced at the cost of decreased performance, but we evaluate using 20MB or less in all our experiments. \rev{We will discuss the oblivious memory costs of each of our data structures and algorithms as we present them}. We show how changes in the oblivious memory budget affect a more complex query in Section~\ref{oblivmem}.

\subsection{Our Guarantees}

Our algorithms leak only the sizes of input, intermediate, and result tables and the physical query plan chosen. This security level is the same as Opaque's oblivious mode~\cite{ZDB+17} and Cipherbase~\cite{cipherbase}. One of our \texttt{SELECT} algorithms also leaks whether the rows returned by a query form a continuous segment of the table queried (e.g. as in a range query), but this algorithm can be turned off if the leakage is deemed too large \rev{and is not used in our performance comparison to prior work.}
\rev{For situations where leaking intermediate table sizes is unacceptable, \name/ also has a padding mode where all intermediate results are padded to a chosen size and query optimization is not applied, leaking nothing about queries but the logical plan and the upper bound on result sizes (like Opaque's padding mode~\cite{ZDB+17}).} \name/ can also be combined with more sophisticated padding techniques, like~\cite{shrinkwrap}, that provide differential privacy instead of full obliviousness to reduce the padding.

\rev{In general, whether the size of intermediate tables is sensitive depends on the application.
For example, in a join of two tables where only one row is selected from each table (say, a customer record and the customer's latest order), the sizes of those intermediate results do not reveal much information; however, a query that selects all of the customer's orders (and then perhaps aggregates them) would let an adversary know how many orders the customer made.
\name/ includes a fused select + project + aggregate operator that can avoid leaking intermediate result sizes even in some multi-operator queries by combining these operations into a single, oblivious operator. 
%Details regarding \name/'s leakage properties appear in Sections~\ref{oblivData}, \ref{oblivOps},  and~\ref{optimizer}. 
}

\rev{Similar to leaking intermediate result sizes,} leaking a query plan can reveal information about the structure of queries, e.g. whether an \texttt{INSERT} or \texttt{JOIN} query was executed, and whether an index was used.
However, \name/ hides query parameters such as \emph{which} key in an index was requested.
For example, by observing the physical plans used, an attacker could learn that a query performed a point lookup on an index, but not which key was requested, or whether the same key is requested again later.
Likewise, \name/'s query planner chooses between different implementations of selection \rev{and join} operators based on the number of matching records, but the attacker does not learn which specific records were chosen (Section~\ref{optimizer}).
In general, there is a fundamental tradeoff between information leakage and performance:
if users want some queries to run faster than others, or to send back a smaller result set, an observer will learn that such a query was executed.
However, in practice, hiding which data was accessed disables many access pattern attacks.

\rev{Finally,} data at rest outside the enclave is encrypted and MACed, and leaks only its size. In both \rev{the padding and no-padding} modes, we do not hide the number of tables in a database or which table(s) a query accesses.
%\rev{
%Although \name/ provides oblivious algorithms for insertion, deletion, and updates, it does not provide full support for transactions.
%Section~\ref{oblivData} discusses concurrency and durability could be built on top of the \name/ engine.}
Beyond hiding data values and access patterns, we make the integrity guarantee that \name/ catches any tampering with data by the malicious OS. We use a series of checks to protect against arbitrary tampering within rows of a table, addition/removal of rows, shuffling of table contents, or rollbacks to a previous system state. %We discuss these protections in Appendix~\ref{integrity}.

Appendix~\ref{formalAppendix} presents a formalization of our security guarantees. We provide security arguments for the obliviousness of each storage method and operator as they appear in the text.

\section{\name/ Architecture and Data Structures}\label{oblivData}
\begin{figure}
\centering
\includegraphics[width=0.6\linewidth]{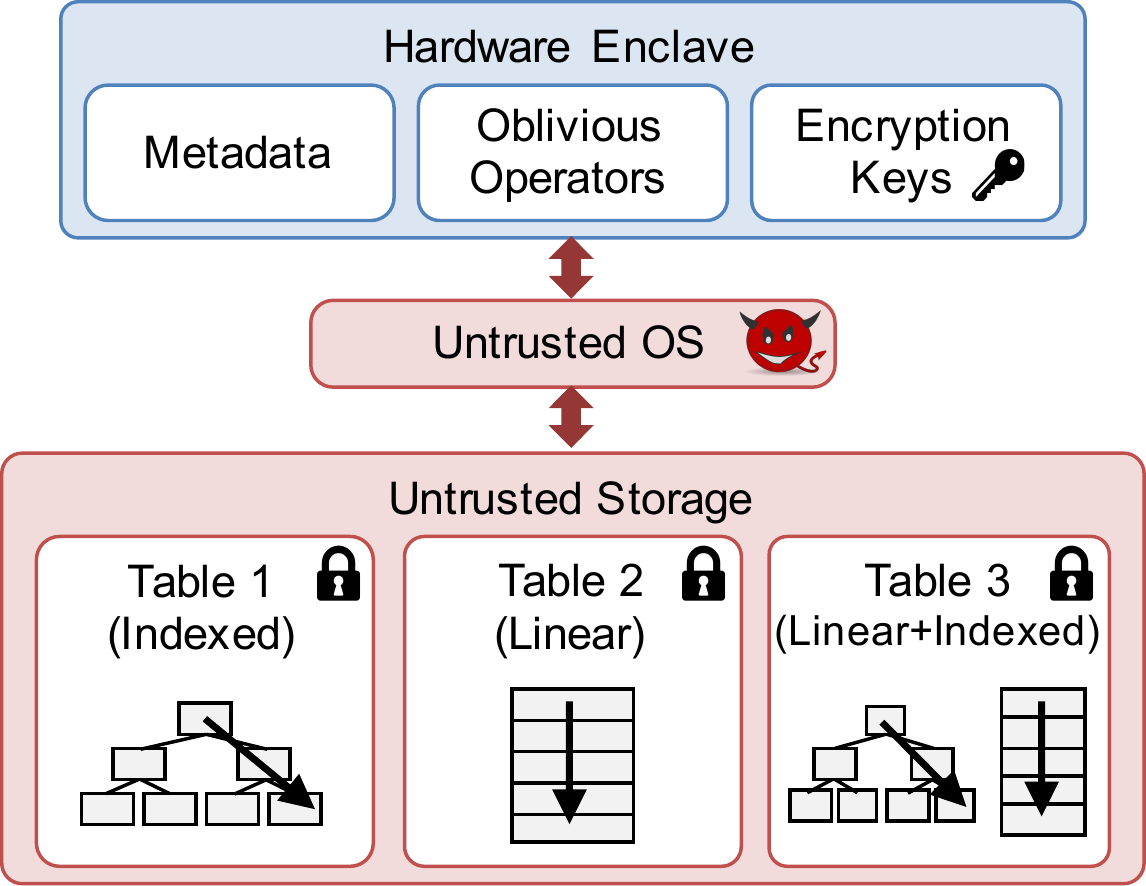}
\caption{\small \name/ runs in a hardware enclave and stores encrypted tables in untrusted memory accessed through the OS. It can store tables using either an oblivious B+ tree index, a flat array, or both.}
\label{arch}
\end{figure}

\begin{figure}
\small
\center
\begin{tabular}{r|ccc}
Method&Flat&Index&Both\\\hline
Space &$N$&$\sim4N$&$\sim5N$\\
Point Read&$O(N)$&$O(\log^2N)$&$O(\log^2N)$\\
Large Read&$O(N)$&$O(N)$&$O(N)$\\
Insert&$O(1)$&$O(\log^2N)$&$O(\log^2N)$\\
Update&$O(N)$&$O(\log^2N)$&$O(N)$\\
Delete&$O(N)$&$O(\log^2N)$&$O(N)$\\
\end{tabular}
\caption{\small Asymptotic performance of storage methods. Fast inserts on flat storage and large reads on indexed storage achieve better than expected asymptotics due to optimizations in Section~\ref{oblivData}.}
\label{asymTables}
\end{figure}

\noindent\textbf{Architecture overview}. Figure~\ref{arch} shows an overview of the \name/ architecture. \name/ consists of a trusted code base inside an enclave that provides an interface for users to create, modify, and query tables using our oblivious query processing algorithms, which we describe in Section~\ref{oblivOps}. \name/ stores tables, authenticated and encrypted, in unprotected memory and obliviously accesses them as needed by the various supported operators. The encryption key for data stored in unprotected memory always resides inside the enclave, encrypting/decrypting blocks of data as they are written or read from unprotected memory. \name/ can store data via two methods: flat and indexed. The indexed method consists of an ORAM with a B+ tree stored inside, whereas the flat method requires scanning the whole table on each query to ensure obliviousness. 

\name/ supports oblivious versions of the operators \texttt{SELECT}, \texttt{INSERT}, \texttt{UPDATE}, \texttt{DELETE}, \texttt{GROUP} \texttt{BY} and \texttt{JOIN} as well as the aggregates \texttt{COUNT}, \texttt{SUM}, \texttt{MIN}, \texttt{MAX}, and \texttt{AVG}. It also includes a query planner that chooses between operator implementations for selection \rev{and join} queries\rev{, which we describe in Section~\ref{optimizer}}.

\rev{Since the core contribution of \name/ lies in its oblivious query processing algorithms, it does not currently include support for transactions, but support for concurrency and logging can be added on top of the current operators. For example, a standard write-ahead log could be generically added to the system. Appends to such a log would not leak any additional information or affect obliviousness, as the only change would be to make a write to an encrypted log file before each insert/update/delete operation. Concurrent access to ORAM data structures could be facilitated by using an ORAM construction that supports parallel access~\cite{obladi,opram1,opram2,opram3,opram4,opram5}. }

\name/ can store data via two methods -- flat and indexed -- or combine both. We currently let system administrators decide which storage method(s) to use for each table based on the expected workload.  Section~\ref{storageCosts} discusses costs and benefits of choosing either or both storage methods. \name/ creates tables with an initial maximum capacity that can be increased later by copying to a new, larger table. We divide data into blocks of a configurable size\footnote{In our current implementation, data in leaves and flat storage are fixed to one record per block.}. Our current implementation assumes records are of fixed length and also stores a boolean flag with each record indicating whether it is in use.

Although encryption and oblivious data structures/algorithms ensure the privacy of data in \name/, additional protections stop an attacker from tampering with data. %Such tampering could take the form of editing within rows of a table, addition/removal of rows, shuffling the contents of a table, or rollbacks to an old system state. \name/ protects against such attacks and reports any attempt to tamper with data.
\name/ MACs and encrypts every block stored outside the enclave, preventing the OS from modifying or adding new rows to tables. Each block of MACed data includes a record of which row(s) the block contains and a current ``revision number'' for that block, a copy of which \name/ also stores inside the enclave. Each time a block is modified, we increment the block's revision number. Any attempt to duplicate, shuffle, or remove rows within a data structure will be caught when an operator discovers that the row number of data it has requested does not exist or does not correspond to that which it has received. Rollbacks of system state are caught when the revision numbers of blocks do not match the last revision numbers for those blocks recorded in the enclave. 
Rollbacks on encrypted enclave data sealed to disk can be prevented either by storing revision numbers with the client or using an enclave rollback protection system like ROTE~\cite{MAK+17}.

\subsection{Flat Storage Method}
The flat storage method simply stores rows in a series of adjacent blocks with no built-in mechanism to ensure obliviousness of memory accesses, so every read or write to the table must involve accesses to every block to hide access patterns. As such, operators acting on these tables, as will be seen in Section~\ref{oblivOps}, involve a series of scans over the entire table. This performs best with small tables, tables where operations will typically require returning large swaths of the table, or analytics that involve reading most or all of the table regardless of the need for obliviousness. The challenge in designing algorithms for this storage method lies in using the limited space of the enclave effectively to reduce the number of scans and data processing operations involved in each operator.

Insertions, updates, and deletions on flat tables involve one pass over the table, during which unaffected blocks receive a dummy write (overwriting a row with the data it already held, re-encrypted and therefore re-randomized). For insertions, the first unused block encountered receives a real write. For updates or deletion, any row matching the specified criteria will be updated or marked unused and overwritten with dummy data, respectively. All of these operations leak nothing about the parameters to the query being executed or the data being operated on except the sizes of the data structures involved because they consist of one scan over a table where each encrypted block is read and then written with a fresh encryption.

In tables with few deletions, an administrator can choose an alternative, constant-time insertion algorithm that saves the index of the last row where an insert occurred and always inserts directly into the next block in memory, skipping the scan. This insertion leaks no additional information beyond the sizes of tables because the access pattern of the insert does not depend on the content of the data except on the number of insertions made, which our adversary can already learn by observing the sizes of tables over time. Since every entry in a table is encrypted, an adversary will not be able to tell if later operations modify or even remove the inserted data, despite knowing ahead of time where each new record will be placed. 

\begin{figure*}
\small
\center
\begin{tabular}{r|llp{7.5cm}}
Algorithm&Time Complexity&\rev{Obliv. Mem.}&Summary\\\hline\rule{0pt}{2.5ex}
Small Select &\rev{$O(N^2/S)$}&\rev{$S$ Bytes}&Fast when data almost fits in enclave: scan table once per enclave-full of data\\
Large Select&$O(N)$&\rev{$0$ Bytes}&Fast when almost entire table selected: copy table and clear unselected rows\\
Cont. Select&$O(N)$&\rev{$0$ Bytes}&Fast when continuous segment of table selected: write to output table for each row of input table, wrapping around at the end, making dummy writes unless row is to be selected\\
Hash Select&$O(N\cdot C)$&\rev{$0$ Bytes}&Use if other strategies don't apply: hash selected rows to location in output table\\
Na\"ive Select&$O(N\log N)$&\rev{$O(R)$ Bytes}&Used only as baseline: ORAM operation for each row of table\\
Aggregate &$O(N)$&\rev{$0$ Bytes}&Scan table, compute aggregate in one pass\\
Gp. Aggregate &$O(N)$&\rev{$O(R)$ Bytes}&Store groups in hash table in oblivious memory and, for each row, check if there is a matching group in the table or add a new group to the table.\\
Hash Join&\rev{$O(\frac{N}{S}\cdot M)$}&\rev{$S$ Bytes}&Block by block, make hash table from one table and see if rows of second table hash to same places -- variant of standard hash join algorithm\\
\rev{Opaque Join}&\rev{$O((N+M)\log^2(\frac{N+M}{S}))$}&\rev{$S$ Bytes}& \rev{Sort tables by join column (use quicksort in obliv. mem. to accelerate), then linear scan to merge blocks of rows. }\\
\rev{0-OM Join}&$\rev{O((N+M)\log^2(N+M))}$&\rev{$0$ Bytes}& \rev{Bitonic sort tables by join column, then linear scan to merge matching rows}\\
\end{tabular}
\caption{\small Oblivious physical operators. $N$ and $M$ are table sizes (in number of rows), $C$ the max chain length of the hash table, \rev{$S$ is the total available oblivious memory, and $R$ the number of rows in the output of a query}. Selection over indexes incurs an additional multiplicative factor of $O(\log^2N)$ in time complexity but runs over the smaller range of rows returned by the index instead of a whole table. \rev{Each indexed table requires $8N$ Bytes of oblivious memory to store and access obliviously.}}
\label{opTable}
%\label{otherOpTable}
\end{figure*}

\subsection{Indexed Storage Method}\label{oblivIndex}
Standard insertion and deletion operations for B+ trees, even when combined with ORAM, leak information about the tree's internal structure, compromising obliviousness by splitting or merging nodes when they reach fixed threshold numbers of children. 
We ensure obliviousness by padding all insertions and deletions with additional dummy ORAM accesses until the number of accesses matches the worst-case number for the respective operation. The property of B+ trees that all data resides in the leaves of the tree means that any lookup already accesses the same number of nodes, so no modification is required for this case. Once each operation involves a fixed number of accesses to memory, we can leverage ORAM's security to guarantee obliviousness. We use the ORAM interface as a black box, so the details of the underlying ORAM can be omitted except to state that our choice of the Path ORAM scheme~\cite{SDS+13} incurs an $O(\log N)$ overhead for each access to memory. See Appendix~\ref{oramAppendix} for details on the construction and formal guarantees of this scheme. \rev{We use a separate ORAM for each table because we already leak which table queries access, and using multiple smaller ORAMs is more computationally efficient than using a single monolithic one.}

Two optimizations dramatically improve the performance of our oblivious B+ trees. First, our implementation operates on a ``lazy write back'' principle, only writing to the ORAM when necessary and otherwise keeping nodes in the enclave until they are no longer needed. 
Second, we remove all parent pointers from our implementation. Normal B+ tree implementations often have pointers in each node to quickly find its parent (e.g.~\cite{BPlus}). 
However, each time a tree splits or merges a node, all the children of nodes involved need to have their parent pointers updated, a very slow process in the regime where every node requires an ORAM write to update. 

If the cost of maintaining both indexed and flat representations of data is too high, e.g. storage is limited or tables are very frequently updated, the indexed storage data structure can also be scanned linearly as a table using the flat method would be, ignoring the index structure. Our algorithms can treat both internal tree nodes and extra ORAM blocks as dummy blocks with no security consequences. This scan has additional overhead over directly using the flat storage method because of the extra space required by ORAM and the index structure, but in practice this overhead is less than 2.5$\times$. 

\subsection{Complexity Analysis}\label{storageCosts}
Figure~\ref{asymTables} compares the asymptotic operations of standard read, insertion, and deletion operations as well as space overhead for each table type. The indexed method performs best on small reads that access one or a few rows of a table, whereas queries which expect to return large segments of a table should use the flat method,  which performs faster than a linear scan over the contents of an index despite equal asymptotic runtimes. Using both storage methods, while incurring the cost of both for insertions and deletions, proves effective when queries of diverse selectivities run on the same data. We empirically measure these tradeoffs in Section~\ref{optimizerEval}. In terms of storage overhead, our oblivious indexes inherit the $4\times$ storage overhead required by the Path ORAM~\cite{SDS+13} we use and each encrypted block is slightly larger than a plaintext block (as is always the case with authenticated encryption). All other sources of storage overhead, e.g. that required for data integrity measures, only add a few bytes to each block of data, amounting to less than $1\%$ additional overhead. \rev{Path ORAM contains a data structure that we need to store in oblivious memory at a cost of 8 Bytes of memory per row of an indexed table. We can reduce the oblivious memory required by using a \emph{recursive} ORAM as described in Appendix~\ref{oramAppendix} or remove it completely via a \emph{doubly-oblivious} ORAM as described by Oblix~\cite{oblix} or ZeroTrace~\cite{SGF17}. }

\section{Oblivious Query Processing}\label{oblivOps}

The key to executing queries in \name/ is a set of new oblivious query processing algorithms that can efficiently run queries over either flat or indexed storage. This section describes our oblivious query processing algorithms for a large subset of SQL, including selection with conditions composed of arbitrary logical combinations of equality or range queries, joins, aggregates (count, sum, max, min, average), and grouped aggregation. In cases where the storage method used for a table admits multiple algorithms to satisfy a given query, \name/'s query planner chooses the algorithm that maximizes performance. At a high level, the planner makes a quick preliminary scan of the table being queried and uses known information about input and output table sizes to make an optimization decision without leaking more information about the query or data. Details on the query planner appear in Section~\ref{optimizer}.

We will begin by discussing the algorithms in the context of flat storage and then discuss the modifications needed for compatibility with indexes, if any. Each operation is accompanied by a security argument. \rev{Since stored rows do not persist inside the enclave between queries, there is no opportunity for a caching side channel based on which rows can be retrieved faster in a subsequent query. Thus the whole engine runs obliviously so long as each of the operators is individually oblivious. }

Whenever we refer to rows of a table being read or written without explicitly stating where they are stored, it is implied that the data resides in unprotected memory, is decrypted before being read inside the enclave, and is re-encrypted before being written back outside. Figure~\ref{opTable} summarizes our algorithms and their complexity. We evaluate the performance of our operators in Section~\ref{eval}.

We refer to the subject of a query as table $T$ and the results as table $R$. We leak only the sizes of $T$ and $R$. In the following, the enclave learns the size of $R$ from the query planner before executing the operator, allowing output data structures of the appropriate size to be allocated before scanning the data needed to fill them.

\subsection{Oblivious Selection Queries}
%\noindent \textbf{Select}.
%Note that satisfying a \texttt{SELECT} query via a straightforward scan that copies each row matching the given criteria into an output table does not provide obliviousness despite touching every row in the table. Such an approach still leaks which rows we include in the output because an attacker can watch the output table and take note of which points in the scan coincide with growth in the output table.
Selection queries involve choosing elements from a table that match a given predicate (e.g. \texttt{date>'2018-09-01'}). One natural way to implement a \texttt{SELECT} operator would be to sequentially read each record in the targeted table and write out the row if it should be selected. Despite touching each row in the table once, this implementation does not provide obliviousness. An adversary observing the pattern of accesses to the input and output tables would know whether a row is written to the output after each read: both tables are accessed each time a row is selected, but only the input table is accessed when the a row is not selected. For example, consider a table \texttt{Checkins} that logs when employees enter or exit an office building. An attacker observing access patterns on the query \texttt{SELECT * FROM Checkins WHERE uid=3172 AND date>'2018-01-01'} could infer from the chosen rows when the user had entered the building or (without seeing the query) what dates the query targets. 

\begin{figure}
\centering
\includegraphics[width=.9\linewidth]{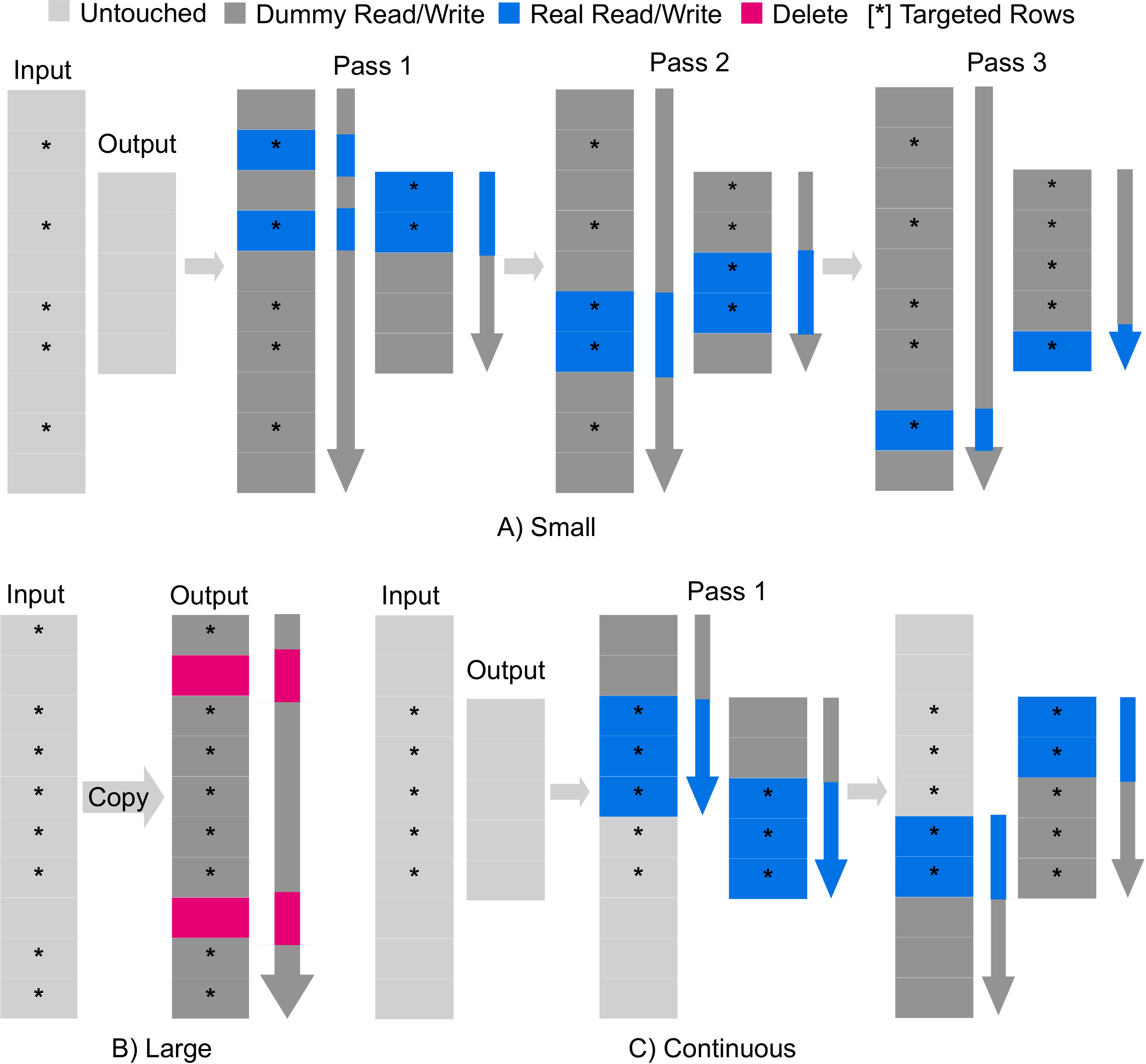}
\caption{\small Small, Large, and Continuous \textsf{SELECT} algorithms. The enclave in this example is only large enough to store two rows of data, so the Small (A) algorithm, which scans the table once per enclave-full of data, takes three passes to complete. The Large (B) and Continuous (C) algorithms always make only one pass. The Large algorithm copies the input table and clears unselected rows, and the Continuous algorithm writes to the output table for each row of the input table, wrapping around at the end, making dummy writes for rows that are not selected.}
\label{selectFigure}
\end{figure}

\begin{figure}
\centering
\includegraphics[width=.9\linewidth]{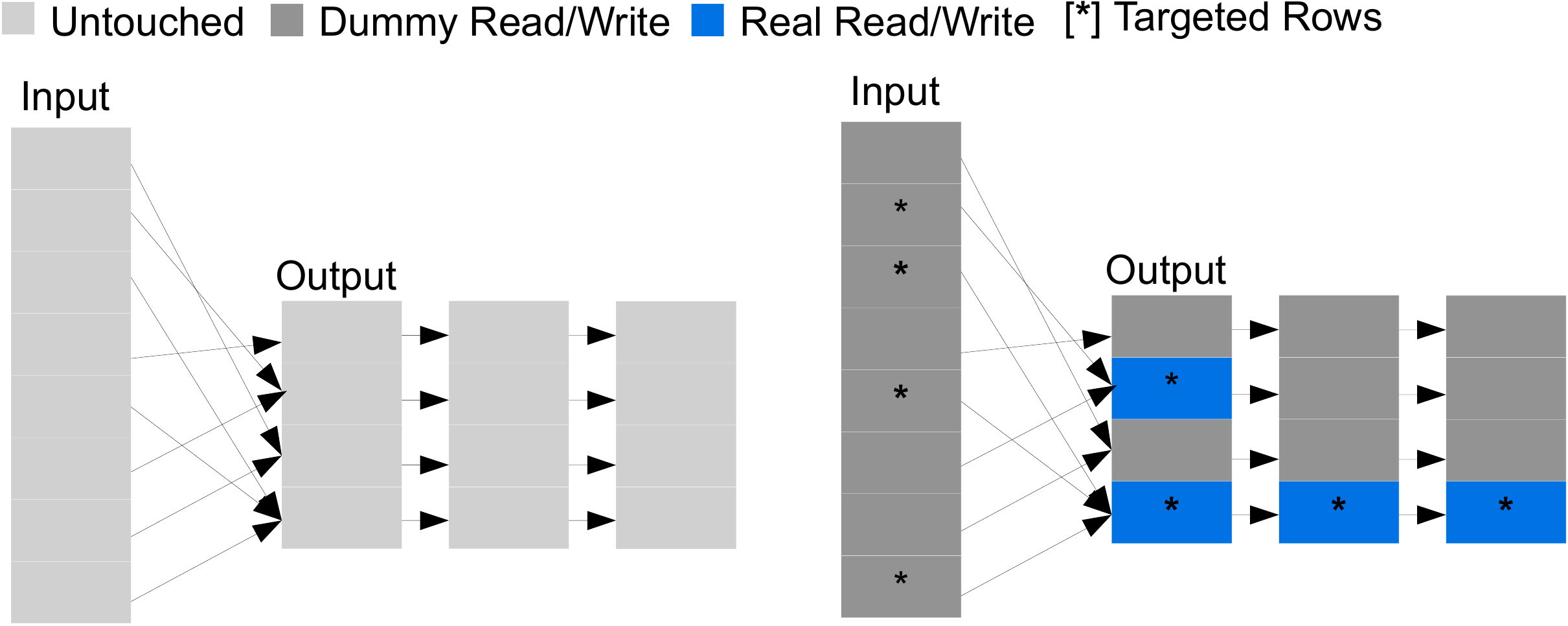}
\caption{\small Hash \textsf{SELECT} algorithm. Left: The access pattern for any input and output table sizes is fixed because the hash of the block number is taken, not of the data itself. Right: a sample execution of the algorithm. Each input cell is read followed by either a dummy or real write following the arrows to the right. Our actual implementation uses double hashing in addition to the chaining shown. }
\label{hashfig}
\end{figure}

To defend against this and other subtle attacks, including those based on prior knowledge of the data distribution, \name/ provides the following oblivious \texttt{SELECT} algorithms (summarized with their complexities in Figures~\ref{opTable},~\ref{selectFigure},~and~\ref{hashfig}). In each algorithm, \name/ has access to the output table size $|R|$ based on information provided by the query planner during its initial scan of the data. 

%\underline{Na\"ive}:
%\subsubsection{Na\"ive}: 
\noindent\textbf{Na\"ive}. 
included as a baseline, the na\"ive oblivious algorithm is a direct translation of a non-oblivious \texttt{SELECT} to an oblivious one via ORAM. After examining each row, it executes an ORAM operation. If the row is included in the output, it makes a write. If not, it makes a dummy read to an arbitrary block. There must be an ORAM operation after reading each row or else an adversary would know that any row which did not coincide with an ORAM operation was not included in the output. After completing the scan of the input table, it copies the contents of the ORAM to the flat storage format and returns it. \rev{This algorithm requires $4|R|$ Bytes of oblivious memory to store the ORAM it uses to build the output table. }

Our techniques to improve on this baseline involve finding the right balance between using data structures in the enclave to remove the need for an ORAM and making multiple fast, oblivious passes over data. These ideas constitute the guiding principle in designing our remaining \texttt{SELECT} algorithms and choosing between them.

%\underline{Small}
%\subsubsection{Small} 
\noindent\textbf{Small}. In the case where all the rows of table $R$ only require a few times the space available in the enclave, a selection strategy that makes multiple fast passes over the data proves effective. We take multiple passes over table $T$, each time storing any selected rows into a buffer in the enclave's oblivious memory and keeping track of the index of the last checked row. Each time the buffer fills, its contents are written to $R$ \emph{after} that pass over $T$. Although this strategy could result in a number of passes linear in the size of $R$, it is effective for small tables. \rev{Since it requires oblivious memory to store rows in the enclave, this algorithm uses whatever quantity of oblivious memory is made available to it. However, reducing the amount of oblivious memory does not affect correctness, only performance. } This algorithm is depicted in Figure~\ref{selectFigure}A.

This algorithm leaks only the sizes of tables $T$ and $R$ because every pass over the data consists of one read to each row and the number of passes reveals only how many times the output set will fill the enclave, a number that can be calculated from the size of $R$. 

%\underline{Large}
%\subsubsection{Large} 
\noindent\textbf{Large}. 
If table $R$ contains almost every row of table $T$, we create $R$ as a copy of $T$ and then make one pass over $R$ where each unselected row is marked unused and each selected row receives a dummy write. Obliviousness holds because the copy operation does not depend on the data copied and we clear unselected rows with a read followed by a write to each block of the table, revealing only the size of $T$. This algorithm, shown in Figure~\ref{selectFigure}B, \rev{ uses no oblivious memory}.

%\underline{Continuous}
%\subsubsection{Continuous}
\noindent\textbf{Continuous}. 
Should the rows selected form one continuous section of the data stored in the table, \name/ requires only one pass over the table, as shown in Figure~\ref{selectFigure}C. Such a situation can arise when range queries are made over sorted data such as names, dates, ID numbers, etc, or retrieved in the same order they were inserted. To handle such queries, \name/ first creates table $R$. Then, for the $i$th row in table $T$, if that row should be in the output, it writes the row to position $i\bmod |R|$ of $R$. If not, it makes a dummy write. Since the rows that need to be included in $R$ make up one continuous segment of $T$, this procedure results in exactly the selected rows appearing in $R$. \rev{This algorithm uses no oblivious memory. }

In addition to the sizes of tables $T$ and $R$, the fact that \name/ chooses this algorithm over one of the other options leaks that the result is drawn from a continuous set of rows in the table. In these cases, however, knowing that users are selecting a range is often not so sensitive as what that range is, which we do hide. Users concerned about this additional leakage could disable this option and use one of the other options with no reduction in supported functionality.  The execution of the algorithm itself is oblivious because the memory access pattern is fixed: at each step, the algorithm reads the next row of $T$ and then writes to the next row of~$R$.

%\underline{Hash}:
%\subsubsection{Hash} 
\noindent\textbf{Hash}. If none of the preceding special-case algorithms apply, \name/ uses a hashing solution illustrated in Figure~\ref{hashfig}. For the $i$th row in $T$, if the row is to be included in the output, we write the content of the row to the $h(i)$th position in $R$, for some hash function $h$. Otherwise, we make a dummy write to the $h(i)$th position in $R$. Since the hash is on the index of the row in the data structure and not over the actual contents of a row, information about the data cannot be leaked by access patterns when rows are written to~$R$\rev{, and the algorithm uses no oblivious memory}.

The algorithm above needs a couple changes to ensure that we properly handle collisions while maintaining obliviousness. We can use standard techniques to resolve collisions, but in order to maintain obliviousness, every row of $T$ must make the same accesses to memory regardless of whether it is included in $R$. We handle this by having every write make as many memory accesses as in the case of the worst expected chain of collisions, regardless of whether the row under consideration in $T$ actually appears in $R$. Following the guidance of Azar et al.~\cite{ABKU99} to get small probability of failure, we use double hashing and have a fixed-depth list of 5 slots for each position in $R$. This means that for each block in $T$, there will be 10 accesses to $R$, 5 for each of the two hash functions. 

The modifications above ensure that data access patterns are fixed regardless of the data in the table and which rows the query selects. As mentioned above, since we hash the index of the row in the data structure and not the actual contents of a row, information about the data itself cannot be leaked by access patterns when rows are written to $R$. As such, we leak only the sizes of $T$ and~$R$.

\noindent\textbf{Selection over Indexes}. Selection over the indexed storage method works identically to flat storage except that the linear scan begins inside an ORAM at a point specified by an index lookup. If the rows returned by a query are not continuous, the leakage also includes the size of the segment of the database scanned in the index. For example, supposing that there is one student named Fred in a table of students indexed by student IDs, the query \texttt{SELECT * FROM students WHERE NAME = ``Fred'' AND ID > 50 and ID < 60} leaks that 9 rows were scanned in the execution of the query. We consider this leakage to be included in the sizes of intermediate tables, as this query is equivalent to a query plan which selects a continuous segment from an index and then selects a noncontinuous segment from the returned table. Padding can hide this leakage. The Large algorithm is not used in the indexed storage method because indexes are meant for queries that request a small fraction of a table, not almost all of it. In terms of complexity, algorithms running over the index have the same complexity as their flat counterparts in Figure~\ref{opTable}, but each algorithm incurs a $\log^2|T|$ multiplicative overhead due to use of the index structure for reads. On the other hand, the actual query runs on table $T'$, the range of rows returned from the query to the index, instead of the full table~$T$. 

\noindent\textbf{Example}. Consider the following queries on a flat table: %recording when employees enter or exit an office building:
\begin{align*}
\small\texttt{SELECT * FROM Checkins WHERE date$=$'2018-08-14'}\\
\small\texttt{SELECT * FROM Checkins WHERE date$>$'1900-01-01'}
\end{align*}
When \name/ receives such queries, it first runs the query planner, which determines which algorithm will perform best for the given query. Since the first query requests rows from a specific date, there will only be a handful of entries, so it chooses the Small algorithm. On the other hand, the second query likely predates the construction of the building and will therefore select every row of the table. The planner will choose the Large algorithm for this query. 

For query 1, \name/ will scan the table, \rev{storing} any records from the chosen date inside the enclave until the end of the scan. Then it will write all the matching rows to an output table at once. If the enclave fills before reaching the end of the table, \name/ will finish the scan without \rev{storing} any more records and then conduct a second scan that begins storing records in the enclave where the first left off. For query 2, \name/ copies the table to create an identical output table and then makes a scan of the copy to delete any rows from before the year 1900. If the above queries were part of a larger query or if the user decided to make a subsequent query on the output, \name/ would then use the output of these queries as the input to the next query and run the appropriate operator. 

\subsection{Oblivious Aggregate \& Group By Queries}

%\noindent \textbf{Aggregates \& Group By}.
An aggregate over a subset or entirety of a table requires only one pass over the table where we calculate the aggregate cumulatively based on the data in each row in $O(|T|)$ time. 
%A na\"ive approach uses ORAM to track the aggregate and needs to update it once for each row, causing an unnecessary slowdown. 
We keep the aggregate statistic inside the enclave. Since the memory access pattern of this operation always involves sequential reads of each block in the table \rev{followed by an update to the aggregate statistic}, nothing leaks beyond the size of table~$T$ \rev{and no oblivious memory is needed}. 

We handle grouped aggregation similarly, except an array in the enclave keeps track of aggregates for each group. \rev{Since we need to hide which group's aggregate each row modifies, we require 4 Bytes of oblivious memory to store the aggregate for each group. } %where a na\"ive solution would check an entire array via ORAM for each row of table $T$. 
We use a hash bucketing approach where each group's value is hashed and inserted into a hash table in the enclave. Each row scanned is hashed and checked against the hash table. If there is a match, then the row under examination corresponds to a known group referenced in the table, and if not, then the current row is added to the hash table as a new group. This method results in running time $O(|T|)$. If the number of groups becomes so large that the hash table cannot fit in oblivious memory (a situation that did not arise in any of our experiments, as each additional group requires very little space), we could switch to using the sort-and-filter approach introduced by Opaque~\cite{ZDB+17} which runs in time $O(|T|\log^2|T|)$.

\noindent\textbf{Combining Aggregation and Selection}. In order to improve performance and avoid leaking intermediate table sizes for common queries that combine selection, aggregation, and grouped aggregation, \name/ provides a combined select/group/aggregate implementation. The \textsf{SELECT} algorithms described above require multiple passes over a table in order to provide obliviousness, but if the next query only takes an aggregate, the obliviously produced intermediate table can immediately be discarded, wasting all the effort of creating it. We remove this inefficiency by computing aggregates directly over the input table while filtering it for the selection criteria. Since selected rows don't need to be written anywhere, we skip the extra effort required by general-purpose oblivious selection.

\subsection{Oblivious Join Queries}\label{oblivjoinqueries}

\rev{Arasu and Kaushik~\cite{AK14} and Opaque~\cite{ZDB+17} previously introduced oblivious join algorithms that are also applicable to \name/. We support Opaque's join algorithm as well as two additional algorithms: an oblivious hash join and a variant of the Opaque join that requires \emph{no oblivious memory}.}

\noindent \textbf{Oblivious Hash Join}. We implement a variant of the standard hash join algorithm~\cite{EN10}. We refer to the two tables being joined as $T_1$ and $T_2$. We make a hash table out of as many rows of $T_1$ as will fit in the enclave and then hash the variable to be joined from each row of $T_2$ to check for matches. This process repeats until reaching the end of $T_1$ . After each check, a row is written to the next block of an output table. If there is a match, the joined row is written. If not, a dummy row is written to the table at that position. Since each comparison between the tables always results in one write to the next block of the output structure, the memory access pattern of this algorithm is oblivious. Like the traditional join algorithm of the same name, the complexity of our oblivious hash join is~$O(|T_1|\cdot|T_2|)$. \rev{Since it needs oblivious memory to store the hash table, this algorithm uses whatever quantity of oblivious memory is made available to it. However, as with the Small selection algorithm, reducing the amount of oblivious memory does not affect correctness, only performance. A side effect of this algorithm's obliviousness is that the size of the output table data structure will always be $|T_1|\cdot|T_2|$. Our remaining join algorithms focus on the case of foreign key joins where the maximum output size is at most the greater of $|T_1|$ and $|T_2|$.}

\noindent \textbf{Oblivious Sort-Merge Join}. \rev{We support two sort-merge join algorithms for foreign key joins. First, we re-implement the Opaque join. This algorithm begins by putting the contents of both tables into one new table. Then it uses quicksort to sort chunks of the data that fit inside an enclave's oblivious memory and merges the chunks with a bitonic sorting network. Finally, one linear scan down the new sorted table eliminates rows that do not have matches and merges matching rows to form the output table.  In addition to requiring oblivious memory, using quicksort to accelerate the join may open timing side channels as well, a factor that must be considered in choosing a join algorithm for a particular application.}

\rev{Next, we support a variant of the Opaque join that requires \emph{no oblivious memory} and operates by running a bitonic sort over the rows of both tables according to the join criteria without quicksorting chunks inside of oblivious memory first. The bitonic sort can be implemented obliviously because it always makes the same set of comparisons independent of the data being sorted. As an optimization, when the size of the recursive sort becomes small enough to fit inside of the enclave, we carry out the sort inside the enclave to avoid paying the cost of calls to memory outside the enclave. This has no impact on obliviousness but speeds up memory access by reducing communication between the enclave and untrusted memory while sorting. We call this the \emph{0-OM} join.}

\rev{We compare the performance of our join algorithms in Section~\ref{eval} and state their complexities in Figure~\ref{opTable}. We could reduce the $O(\log^2n)$ terms in the oblivious sorts to  $O(\log n)$ using a randomized shellsort~\cite{Goodrich11} (as discussed by Arasu and Kaushik~\cite{AK14}) at the cost of making the correctness of the sorting algorithm probabilistic. }

\section{Query Planner}\label{optimizer}

Our query planner picks which selection \rev{and join} algorithms to use based on statistical information on the \rev{input and output table sizes}. Our main insight is that we can use the information already leaked by the data structures and output sizes in \name/ to minimize additional leakage from query planning. In Section~\ref{optimizerEval}, we find that the planner can improve query performance by $4.6$-$11\times$. The query planner is not used in padding mode, where we hide output sizes.

\name/ runs the query planner at runtime whenever it encounters a selection \rev{or join} operator. For each selection, the planner begins with a fast scan over the data, during which it keeps track of (1) the number of rows satisfying the predicate and (2) whether those rows are adjacent in the input table. The enclave saves the computed output size to pass into selection operators that pre-allocate output storage. Based on the ratios of number of output rows to available \rev{oblivious memory} and input table size, the planner decides which variant of the selection operator to use. A precomputed set of thresholds decide when to run each operator. For maximum flexibility, users can also manually choose to force a particular operator.

Note that we cannot simply return the query result in the first scan over the data, as a na\"ive one-pass algorithm would violate obliviousness. Instead, we must run one of our oblivious operators. Since many of these operators need to know the size of the output table up-front (to allocate memory for the results), the planner's first scan to compute statistics is often ``for free.'' 

\rev{We adopt a similar approach to choose the appropriate algorithm for foreign key joins, but planning for joins requires even less information than selection. Observe that all the join algorithms in Section~\ref{oblivjoinqueries} generate output tables and do computation of the maximum possible size given the input table sizes. As such, the output table, although it may contain many dummy rows that are marked as unused, will reside in a data structure whose size can be calculated directly from the sizes of the input tables. Moreover, this property means that the performance of the join algorithms will depend only on the input table sizes and will otherwise be the same \emph{regardless of the selectivity} of the join. These properties taken together allow us to make effective optimization decisions based only on knowledge of the sizes of the tables joined and the amount of oblivious memory available inside the enclave. Similar to selection, we pick which join algorithm to use based on the ratio of the available oblivious memory to the size of the first input table. If the amount of oblivious memory is large relative to the size of the first table, we always use the hash join. Otherwise, we plug in the table sizes and amount of oblivious memory into expressions denoting the asymptotic runtimes of the join algorithms and choose the smaller result. Section~\ref{optimizerEval} shows that this approach works well in practice.}

\noindent\textbf{Security}. Performance improvements due to query planning intrinsically require leakage because the benefits of planning arise from the fact that different algorithms perform better for different data and queries. Our choice of physical operator reveals two pieces of information. First\rev{, for selection,} is the number of matching rows. Since the non-padded execution mode already reveals the output size of the result, this adds nothing to the overall leakage of the system. Second is whether or not the rows returned by a query form a continuous segment of the table queried. This is revealed by the choice of the Continuous algorithm from Section~\ref{oblivOps}, which occurs if the rows to be returned are continuous. The Continuous algorithm can optionally be disabled, causing optimization to leak \emph{no additional information} beyond what is already revealed through output sizes \rev{(this is the configuration used for our comparison to prior work in Section~\ref{eval}). Planning for joins leaks even less, as it relies only on the sizes of the tables being joined and the oblivious memory available}.

Our query planner always has the same memory access pattern \rev{for selection queries}: read each row, update statistics, and perform a table lookup to select an algorithm at the end. As such, the only leakage introduced by the query planner comes from its final choice of which physical operator to run, not the optimization algorithm itself. \rev{For joins, the planner only reads the recorded sizes of the input tables and makes no other memory accesses.}

\section{Implementation}\label{imp}
We implemented \name/ on Intel SGX~\cite{CD16}, including the storage methods from Section~\ref{oblivData} as well as the oblivious operator algorithms and query planner of Sections~\ref{oblivOps} and~\ref{optimizer}. Our implementation consists of over 14,000 lines of code and builds upon the Remote Attestation sample code provided with the SGX SDK ~\cite{SGXRef} and the B+ tree implementation of~\cite{BPlus}, the latter of which was heavily edited in order to support our ORAM memory allocator. We use SGX SDK libraries for encryption, MACs, and hashing. We instantiate our ORAM scheme with a nonrecursive Path ORAM~\cite{SDS+13}. See Appendix~\ref{oramAppendix} for details on this scheme and the oblivious storage and performance implications of recursive vs nonrecursive Path ORAM.

Our current implementation consists only of the core database engine and lacks some components of a full-featured DBMS, e.g. transaction management and persistence to disk. In our evaluation, we compare \name/ only to in-memory tables on other oblivious systems to avoid giving it an unfair advantage. It would be straightforward to replace \name/'s external memory with disk storage, as accesses to both ORAM and flat tables are already block-oriented. \rev{We discuss options for supporting transactions in Section~\ref{oblivData}}.

\begin{figure}
\small
\centering
\begin{tabular}{llp{4cm}}
\textbf{Table Name} & \textbf{Rows} & \textbf{Notes} \\\hline\rule{0pt}{2ex}
%\small CFPB & \small 107,000 & \small complaints to the US Consumer Financial Protection Bureau~\cite{CFPB}.\\\small 
 USERVISITS & 350,000 & Server logs for many sites. Data from the Big Data Benchmark~\cite{BDB}.\\\rule{0pt}{2ex}
RANKINGS & \small 360,000 & URLs, PageRanks, and average visit durations for many sites. Data from the Big Data Benchmark~\cite{BDB}.\\
\end{tabular}
\caption{\small Data sets used in the Big Data Benchmark~\cite{BDB}.}
\label{tabletable}
\end{figure}

\begin{figure*}
\begin{minipage}{.64\textwidth}
\small
\centering
\begin{tikzpicture}
\centering
\begin{semilogyaxis}[
	log origin = infty,
	width=5cm,height=3.3cm,
	y label style={at={(axis description cs:-.2,.5)},anchor=south},
	ytick={.1,1,10},
	ymin=.1,
	ylabel={Time [seconds]},
    title style={align=center},title={Comparison to Opaque\\No Index},
	symbolic x coords={Q1, Q2, Q3, 4},
	x tick label style={
		/pgf/number format/1000 sep=},
	%%ylabel=Time (seconds),
	enlargelimits=0.05,
	legend style={at={(0.5,-0.34)},
	legend entries={\small Opaque Oblivious Mode, \small \name/ (without Index), \small Spark SQL (No Security)},
	anchor=north,legend columns=1},
	ybar interval=.7,
    ]
%\addplot %baseline
%	coordinates {(BDB Q1, 54.18978) (BDB Q2, 205.78637) (BDB Q3, 131.29887) (4,1)};
\addplot %opaque obliv
	coordinates {(Q1,1.54659) (Q2,3.210607) (Q3,17.26355) (4, 1)};
\addplot %us
	coordinates {(Q1,2.434) (Q2, 3.0422) (Q3,12.774) (4,1)};
%\addplot %spark sql
%	coordinates {(BDB Q1,.1615) (BDB Q2,1.9757) (BDB Q3,7.2477) (4,1)};
    %\legend{Opaque Oblivious Mode,\name/, Spark SQL (No Security)}

\addlegendimage{line legend,black,sharp plot,dashed}
\coordinate (A) at (axis cs:Q1,.11755);
\coordinate (B) at (axis cs:Q2,.11755);
\coordinate (C) at (axis cs:Q2,1.2848);
\coordinate (D) at (axis cs:Q3,1.2848);
\coordinate (E) at (axis cs:Q3,4.8357);
\coordinate (F) at (axis cs:4,4.8357);

\draw [black,sharp plot,dashed] (A) -- (B);
\draw [black,sharp plot,dashed] (C) -- (D);
\draw [black,sharp plot,dashed] (E) -- (F);

\end{semilogyaxis}
\end{tikzpicture}
\begin{tikzpicture}
\centering
\begin{semilogyaxis}[
	log origin = infty,
	width=5cm, height=3.3cm,
	y label style={at={(axis description cs:-.2,.5)},anchor=south},
	ylabel={Time [seconds]},
    title style={align=center},title={Comparison to Opaque\\Index Allowed},
	symbolic x coords={Q1,  Q2,  Q3, 4},
	x tick label style={
		/pgf/number format/1000 sep=},
	%%ylabel=Time (seconds),
	enlargelimits=0.05,
	legend style={at={(0.5,-0.34)},
		legend entries={\small Opaque Oblivious Mode, \small \name/ (Index Allowed), \small Spark SQL (No Security), \name/},
	anchor=north,legend columns=1},
	ybar interval=.7,
    ]
%\addplot %baseline
%	coordinates {( Q1, 54.18978) ( Q2, 205.78637) ( Q3, 131.29887) (4,1)};
\addplot %opaque obliv
	coordinates {(Q1,1.54659) (Q2,3.210607) (Q3,17.26355) (4, 1)};
\addplot %us
	coordinates {(Q1,.0824) (Q2, 3.0422) (Q3,12.774) (4,1)};
%\addplot %spark sql
%	coordinates {(Q1,.1615) (Q2,1.9757) (Q3,7.2477) (4,1)};
    %\legend{Opaque Oblivious Mode,\name/, Spark SQL (No Security)}

    \addlegendimage{line legend,black,sharp plot,dashed}
\coordinate (A) at (axis cs: Q1,.11755);
\coordinate (B) at (axis cs: Q2,.11755);
\coordinate (C) at (axis cs: Q2,1.2848);
\coordinate (D) at (axis cs: Q3,1.2848);
\coordinate (E) at (axis cs: Q3,4.8357);
\coordinate (F) at (axis cs:4,4.8357);
%\coordinate (G) at (60,-2);
%\coordinate (H) at (90,-2);
%\coordinate (I) at (60,-4);
%\coordinate (J) at (90,-4);

\draw [black,sharp plot,dashed] (A) -- (B);
\draw [black,sharp plot,dashed] (C) -- (D);
\draw [black,sharp plot,dashed] (E) -- (F);
%\draw [red,sharp plot,solid] (G) -- (H);
\end{semilogyaxis}
\end{tikzpicture}
\end{minipage}
\begin{minipage}{.35\textwidth}
\small
\begin{tikzpicture}
\begin{axis}[
    title style={align=center}, title={Effect of Oblivious Memory Budget\\Big Data Benchmark Q3},
    width=5cm, height=3.5cm,
    xlabel={Oblivious Memory (MB)},
	y label style={at={(axis description cs:-.1,.5)},anchor=south},
	ylabel={Time [seconds]},
    xmin=4, xmax=22,
    ymin=0, ymax=35,
    xtick={4,8,12,16,20},
    legend pos=south east,
    ymajorgrids=true,
    grid style=dashed,
    	legend style={at={(0.5,-.48)},
		legend entries={\small Opaque, \small \name/},
	anchor=north,legend columns=2},
    ]
    \addplot[
    color=blue,
    mark=square,
    ]
    coordinates {
    (6,31.3164)(8,30.4219)(10,30.3000)(12,28.4505)(14,28.5447)(16,26.8832)(18,26.66089)(20,27.43052)
    };
    \addplot[
    color=red,
    mark=square,
    ]
    coordinates {
    (6,23.46747)(8,23.45914)(10,18.60779)(12,18.59869)(14,18.57459)(16,18.49345)(18,13.55271)(20,13.42186)
    };
    %\legend{\scriptsize Sophos, \scriptsize \name/}
\end{axis}
\end{tikzpicture}\vspace{.3cm}
\end{minipage}
\begin{minipage}{.64\textwidth}
\caption{\small \name/ outperforms Opaque Oblivious~\cite{ZDB+17} by 1.1-19$\times$ and never runs more than 2.6$\times$ slower than Spark SQL~\cite{SparkSQL} on Queries Q1-Q3 of the Big Data Benchmark~\cite{BDB}. Even without use of an index, \name/ performs comparably to Opaque Oblivious.}
\label{figOpaque}
\end{minipage}\hspace{.2cm}
\begin{minipage}{.32\textwidth}
\caption{\small Performance of \name/ and Opaque~\cite{ZDB+17} on Big Data Benchmark Query~3 as oblivious memory varies.}
\label{figoblivmem}
\end{minipage}
%\end{figure*}
%\begin{figure*}
        \centering
\begin{tikzpicture}
\begin{semilogxaxis}[
    title={Retrieval (One Entry)},
        ylabel={Time [ms]},
        	width=5.5cm, height=4cm,
    xlabel={Size of Table (Rows)},
    xtick={100,1000,10000,100000,1000000},
    legend pos=north west,
    ymajorgrids=true,
    grid style=dashed,
    ]
    \addplot[
    color=blue,
    mark=square,
    ]
    coordinates {
    (100,2.75)(1000, 6.57)(10000, 10.99)(100000,19.55)(1000000,27.29)
    };
    \addplot[
    color=red,
    mark=square,
    ]
    coordinates {
    (100,1.28)(1000, 1.54)(10000, 2.19)(100000,2.86)(1000000,3.58)
    };
    \addplot[
    color=orange,
    mark=square,
    ]
    coordinates {
    (100,.324)(1000, .342)(10000, .342)(100000,.317)(1000000,.327)
    };
    \legend{\small HIRB, \small \name/, \small MySQL}
\end{semilogxaxis}
\end{tikzpicture}
\begin{tikzpicture}
\begin{semilogxaxis}[
    title={Insertion},
        ylabel={Time [ms]},
        	width=5.5cm, height=4cm,
    xlabel={Size of Table (Rows)},
    xtick={100,1000,10000,100000,1000000},
    legend pos=north west,
    ymajorgrids=true,
    grid style=dashed,
    ]
    \addplot[
    color=blue,
    mark=square,
    ]
    coordinates {
    (100,2.67)(1000, 6.37)(10000, 10.86)(100000,19.34)(1000000,28.48)
    };
    \addplot[
    color=red,
    mark=square,
    ]
    coordinates {
    (100,.69)(1000, 1.68)(10000, 2.16)(100000,4.25)(1000000,8.29)
    };
    \legend{\small HIRB, \small \name/}
\end{semilogxaxis}
\end{tikzpicture}
\begin{tikzpicture}
\begin{semilogxaxis}[
    title={Deletion},
        ylabel={Time [ms]},
        	width=5.5cm, height=4cm,
    xlabel={Size of Table (Rows)},
    xtick={100,1000,10000,100000,1000000},
    legend pos=north west,
    ymajorgrids=true,
    grid style=dashed,
    ]
    \addplot[
    color=blue,
    mark=square,
    ]
    coordinates {
    (100,2.35)(1000, 6.48)(10000, 11.25)(100000,19.07)(1000000,28.25)
    };
    \addplot[
    color=red,
    mark=square,
    ]
    coordinates {
    (100,1.36)(1000, 2.26)(10000, 3.42)(100000,6.19)(1000000,9.36)
    };
    \legend{\small HIRB, \small \name/}
\end{semilogxaxis}
\end{tikzpicture}
    		\caption{\small \name/'s oblivious indexes outperform the HIRB tree + vORAM oblivious map construction.}
    		\label{HIRB}
%\end{figure*}
\ignore{
\begin{figure}
\centering
\begin{tikzpicture}
\begin{loglogaxis}[
    title style={align=center}, title={Comparison to Sophos\\Select from 1.4 Million Row Table},
    width=5cm, height=3.5cm,
    xlabel={Rows Selected},
	y label style={at={(axis description cs:.1,.5)},anchor=south},
	ylabel={Time [seconds]},
    xmin=5, xmax=15000,
    ymin=0, ymax=130,
    xtick={10,100,1000,10000},
    legend pos=south east,
    ymajorgrids=true,
    grid style=dashed,
    	legend style={at={(0.5,-.48)},
		legend entries={\small Sophos, \small \name/},
	anchor=north,legend columns=2},
    ]
    \addplot[
    color=blue,
    mark=square,
    ]
    coordinates {
    (10,.2)(100,.8)(1000,7)(10000,69)
    };
    \addplot[
    color=red,
    mark=square,
    ]
    coordinates {
    (10,.00885)(100,.07083)(1000,.69767)(10000,10.24335)
    };
    %\legend{\scriptsize Sophos, \scriptsize \name/}
\end{loglogaxis}
\end{tikzpicture}
\caption{\small Comparison to Sophos SSE scheme\cite{Bost16}. \name/ always outperforms Sophos by at least 22.6$\times$.}
\label{figSophos}
\end{figure}
}
%\begin{figure*}
\vspace{1em}
\small
\centering
\begin{tikzpicture}
\begin{axis}[
    title={Flat vs Index Select},
    	width=5cm, height=2.8cm,
    xlabel={Percent of Table Retrieved},
    ylabel={Time [seconds]},
    xmin=0, xmax=2.5,
    ymin=0, ymax=1.5,
    xtick={0,.5,1,1.5,2, 2.5},
	legend style={at={(0.5,-0.75)},
	anchor=north,legend columns=-1},
    ymajorgrids=true,
    grid style=dashed,
    ]
    \addplot[
    color=blue,
    mark=square,
    ]
    coordinates {
    (.1,.6886)(.5,.6784)(1,.6968)(1.5,.7203)(2,.6922)(2.5,.7049)
    };
    \addplot[
    color=red,
    mark=square,
    ]
    coordinates {
    (.1,.0556)(.5,.2732)(1,.5666)(1.5,.8177)(2,1.0866)(2.5,1.3544)
    };
    \legend{\small Flat, \small Indexed}
\end{axis}
\end{tikzpicture}
\begin{tikzpicture}
\begin{axis}[
    title={Flat vs Index Group By},
        ylabel={Time [seconds]},
        	width=5cm,height=2.8cm,
    xlabel={Percent of Table Retrieved},
    xmin=0, xmax=2.5,
    ymin=0, ymax=.75,
    xtick={0,.5,1,1.5,2, 2.5},
    ytick={0,.25,.5,.75},
	legend style={at={(0.5,-0.75)},
	anchor=north,legend columns=-1},
    ymajorgrids=true,
    grid style=dashed,
    ]
    \addplot[
    color=blue,
    mark=square,
    ]
    coordinates {
    (.1,.3462)(.5,.3393)(1,.3495)(1.5,.3546)(2,.3458)(2.5,.3465)
    };
    \addplot[
    color=red,
    mark=square,
    ]
    coordinates {
    (.1,.0282)(.5,.1361)(1,.2816)(1.5,.4025)(2,.5365)(2.5,.6744)
    };
    \legend{\small Flat, \small Indexed}
\end{axis}
\end{tikzpicture}
\begin{tikzpicture}
\begin{semilogyaxis}[
	log origin = infty,
	width=4.7cm,height=3.07cm,
    title={Flat vs Index Operations},
        ylabel={Time [ms]},
	symbolic x coords={Insert, Delete, Update, 4},
	x tick label style={
		/pgf/number format/1000 sep=},
	%%ylabel=Time (seconds),
	enlargelimits=0.05,
	legend style={at={(0.5,-0.37)},
	anchor=north,legend columns=-1},
	ybar interval=0.7,
    ]
\addplot
	coordinates {(Insert,.01) (Delete,706.34) (Update,710.77) (4, 1)};
\addplot
	coordinates {(Insert,004.24) (Delete,005.93) (Update,002.68) (4,1)};
    \legend{Flat, Indexed}
\end{semilogyaxis}
\end{tikzpicture}
\caption{\small Comparison of flat and indexed versions of operators over 100,000 rows of synthetic data. Flat scans do better when more data needs to be accessed, but the indexed storage method performs far better for small queries.}
\label{figC1}\vspace{.5cm}
    \centering
    \begin{minipage}{.32\textwidth}
        \centering
\begin{tikzpicture}
\begin{semilogxaxis}[
    title={Point Queries on Indexes},
        ylabel={Time [ms]},
        	width=5.5cm, height=4cm,
    xlabel={Size of Table (Rows)},
    xtick={100,1000,10000,100000,1000000},
    legend pos=north west,
    ymajorgrids=true,
    grid style=dashed,
    ]
    \addplot[
    color=blue,
    mark=square,
    ]
    coordinates {
    (100,1.28)(1000, 1.54)(10000, 2.19)(100000,2.86)(1000000,3.58)
    };
    \addplot[
    color=red,
    mark=square,
    ]
    coordinates {
    (100,.69)(1000, 1.68)(10000, 2.16)(100000,4.25)(1000000,8.29)
    };
    \addplot[
    color=orange,
    mark=square,
    ]
    coordinates {
    (100,1.36)(1000, 2.26)(10000, 3.42)(100000,6.19)(1000000,9.36)
    };
    \legend{\small \texttt{SELECT},\small \texttt{INSERT},\small \texttt{DELETE}}
\end{semilogxaxis}
\end{tikzpicture}
    \end{minipage}
    \hspace{.2cm}%
    \begin{minipage}{0.64\textwidth}
    \begin{minipage}{.5\textwidth}
\begin{tikzpicture}
\centering
\begin{semilogyaxis}[
	log origin = infty,
	width=5.15cm, height=3.85cm,
	ybar interval=.7,
	ylabel={Ops/Second},
    title={Impact of Table Type},
	symbolic x coords={\ \ \ L1, \ \ \ L2,  \ \ \ L3,  \ \ \ L4,  \ \ \ L5, 4},
	legend style={at={(0.5,-.25)},
	anchor=north,legend columns=4},
	xtick=data,
	x tick label style={text width=1cm},
	%%ylabel=Time (seconds),
	enlargelimits=0.05,
    ]
\addplot
	coordinates {(\ \ \ L1, 14.401) (\ \ \ L2, 1.622) (\ \ \ L3, 1.424) (\ \ \ L4, 1.499) (\ \ \ L5, 1.486) (4,1)};
\addplot
	coordinates {(\ \ \ L1, 7.081) (\ \ \ L2, 37.539) (\ \ \ L3, .725) (\ \ \ L4, .806) (\ \ \ L5, .408) (4,1)};
\addplot[color=violet, fill=violet]
	coordinates {(\ \ \ L1, 25.240) (\ \ \ L2, 29.682) (\ \ \ L3, 2.765) (\ \ \ L4, 2.768) (\ \ \ L5, 1.485) (4,1)};

\legend{\small Flat, \small Indexed, \small Both}
\end{semilogyaxis}
\end{tikzpicture}
\end{minipage}
      \begin{minipage}{.5\textwidth}\vspace{.2cm}
\centering
\small
\setlength\tabcolsep{3.5pt}
\begin{tabular}{llllll}
\textbf{Workload}&L1&L2&L3&L4& L5\\\hline
\% Point Reads & 5& 0& 50& 45& 0\\
\% Small Reads & 0& 90& 0& 0& 0\\
\% Large Reads & 5&0&50&45&90\\
\% Insertions& 90& 9& 0& 5& 5\\
\% Deletions & 0& 1& 0& 5& 5\\
\end{tabular}
\end{minipage}
    \end{minipage}
    \begin{minipage}{.32\textwidth}
    		\caption{\small Point queries for various table sizes. Query time is polylogarithmic in table size.}
		\label{figScaling}
    \end{minipage}\hspace{.2cm}
    \begin{minipage}{.64\textwidth}
    \caption{\small Flat, indexed, and combined representations of a 100,000 row table for five workloads. Point reads access 1 row, small reads access 50, and large reads access 5\% of the table.}
\label{figworkload}
    \end{minipage}
\end{figure*}

\section{Evaluation}\label{eval}

We evaluate \name/ on multiple datasets, comparing to prior private database systems and widely used non-private systems. We use a subset of the data available from the Big Data Benchmark~\cite{BDB}, shown in Figure~\ref{tabletable}, as well as larger synthetic data. In addition, we measure the overhead of \name/'s padding mode,  demonstrate the effectiveness of \name/'s query planner, study the impact of the chosen storage methods, and examine tradeoffs in join algorithms through a series of microbenchmarks. We evaluated \name/ on an Intel Core i7-6700 CPU @3.4GHz with 8GB of RAM running Ubuntu 16.04 and the SGX SDK version 1.9. \rev{The comparison of join algorithms was done on the same machine running Ubuntu 18.04 and the SGX SDK version 2.5.}

We find that \name/ can leverage its indexes to achieve order of magnitude performance improvements over previous private database systems. In particular, \name/ matches Opaque~\cite{ZDB+17} for scan-based queries on flat tables but can outperform it by 19$\times$ when using an index. \name/ also performs over 7$\times$ faster than HIRB~\cite{RAC16}, an oblivious map scheme, and comes within a factor of $2.6\times$ the performance of the non-private Spark SQL system.

\subsection{Comparison to Prior Work}
\noindent\textbf{Comparison to Opaque}. Figure~\ref{figOpaque} compares \name/ with Opaque's oblivious mode \cite{ZDB+17,opaquecode} and Spark SQL~\cite{SparkSQL}, which provides no security guarantees, on queries 1-3 of the Big Data Benchmark~\cite{BDB} on tables of 360,000 and 350,000 rows. \rev{We use the same queries and parameters used by Opaque: 1000, 8, and 1980-04-01 are the parameters used for queries 1-3 of the benchmark, respectively. Query 1 targets straightforward selection, Query 2 tests grouped aggregation, and Query 3 tests joins.} Opaque also uses an SGX enclave and can be configured in either ``encryption'' mode or ``oblivious'' mode, which hides access patterns to data, but by means different from ours. We compare to Opaque's oblivious mode and run it in single node configuration. We limit oblivious memory to 72MB for Opaque (as in its original evaluation) and 20MB for \name/, but neither system needed the full oblivious memory allowed. \rev{To compare fairly in terms of the leakage permitted, we disable Continuous selection algorithm in the comparison with Opaque.}

We began by configuring \name/ to use only the flat storage method, as Opaque does, and found that \name/ performs comparably to Opaque, slightly worse on query 1 and slightly better on queries 2 and 3. Next, we used the combined storage method. An oblivious index allows \name/ to outperform Opaque by 19$\times$ on query 1 since this query scans a small part of a table whereas Opaque and spark SQL, which primarily handle analytic workloads, scan the entire table. Indexes do not provide a speedup on queries 2 and 3 which scan most of the input anyway. \name/ is only 2.4$\times$ and 2.6$\times$ slower than Spark SQL on queries 2 and~3.

We also tested scan-based queries against our indexes to see how \name/ performs on frequently-updated data too expensive to maintain in flat storage.
These queries performed about \textbf{2$\times$} slower than on flat tables.
Thus, unlike prior, flat-only systems, \name/ performs analytics relatively quickly on ``live'' tables frequently updated with point insertions and deletions.

\noindent\textbf{Impact of Oblivious Memory Budget}.\label{oblivmem}
Figure~\ref{figoblivmem} shows the performance of \name/ and Opaque's oblivious mode on query~3 of the Big Data Benchmark as the quantity of oblivious memory varies from 6MB to 20MB, beyond which the performance of \name/ remains steady. We chose this query because its performance is most affected by an increase in oblivious memory for both systems. Both systems' performance improves as we add more oblivious memory, but Opaque improves gradually whereas \name/ decreases in steps as the amount of oblivious memory makes the blocks of the nested loop join large enough to reduce the overall number of scans of the second table being joined. In total, the increase from 6MB oblivious memory to 20MB results in a $1.77\times$ speedup for \name/. 

\noindent\textbf{Comparison to HIRB}.\label{HIRBcomp}
Next, we compare \name/'s performance to The HIRB Tree + vORAM~\cite{RAC16} secure index structure. Unlike \name/, HIRB neither supports range queries nor uses hardware enclaves. Source code for other SGX-based oblivious indexes is not yet publicly available, so we cannot compare to them directly, although reported numbers for Oblix~\cite{oblix} and POSUP~\cite{POSUP} appear to be within several milliseconds of ours. Despite its reduced functionality and differing security assumptions, HIRB provides a good point of comparison as a practical system attempting to solve similar problems. We compare against it with a replication of the performance experiment in its original paper. 

%\textbf{HIRB Tree + vORAM.}
Figure~\ref{HIRB} compares the point query performance of \name/'s oblivious indexes with a HIRB tree + vORAM oblivious map~\cite{RAC16} and MySQL. \rev{Although \name/ does not support transactions, we include comparisons of insertion and deletion times over our indexes to demonstrate the performance of the data structure (the comparison is fair since HIRB also implements a key-value store with no notion of concurrency or durability).} We instantiated both the table in \name/ and the HIRB tree with 64-Byte data entries and allocated the underlying vORAM with bucket size 4096, a somewhat larger size than our own ORAM's buckets (HIRB performed worse on smaller bucket sizes). On tables of 1,000,000 rows, \name/ outperforms HIRB by 7.6$\times$ in point selection and by 3$\times$ on insertions and deletions. While still an order of magnitude slower than MySQL for \rev{point queries on} larger tables, network latency from user to cloud can be tens of milliseconds, rendering the difference insignificant. 

The HIRB construction considers a ``catastrophic attack'' scenario which compromises the system holding the ORAM client, and they design the HIRB tree to provide history independence and secure deletion even under this attack. Since our work relies on the security of the hardware enclave and keeps the ORAM client inside the enclave, the additional security properties desired by HIRB come for free in our setting, explaining our improved performance. Both our work and the HIRB tree make use of padding for obliviousness, but each uses different optimizations to minimize padding.

\ignore{
\textbf{Sophos.}
We compare \name/'s oblivious index to the searchable symmetric encryption (SSE) scheme Sophos~\cite{Bost16} in Figure~\ref{figSophos}. Sophos does not provide obliviousness guarantees, meaning it leaks access patterns. It does provide a good point of comparison for the performance of our SGX-based oblivious indexes with a non-SGX based, non-oblivious index that still provides some privacy. We compare against numbers reported in the Sophos paper for a 1.4 million row table using a more powerful machine than ours: an Intel Core i7 4790K 4.00GHz CPU with 8 logical cores and 16GB of RAM running on OS X.10. Despite the difference in hardware and the fact the Sophos is multithreaded, \name/ outperforms Sophos by 22.6-24.6$\times$. We observe that the performance tipping point between indexed and flat storage methods in this experiment arrives between $10^4$ and $10^5$ rows, and \name/'s performance on larger queries beyond that point would remain constant. \name/ performs better than Sophos because the tens of AES invocations needed for ORAM accesses are much cheaper than Sophos's public-key cryptography.
}

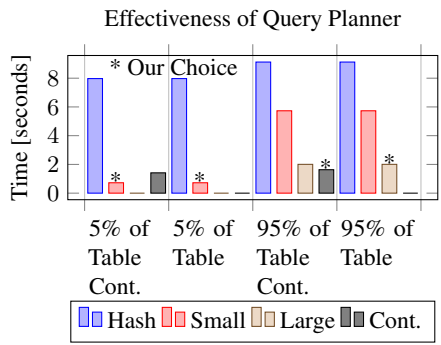
\begin{figure}
    \small
    \centering
    \begin{tikzpicture}
    \centering
    \begin{axis}[
        width=6.5cm, height=3.5cm,
        ybar interval=.7,
        ylabel={Time [seconds]},
        title={Effectiveness of Query Planner},
        symbolic x coords={5\% of Table Cont., 5\% of Table, 95\% of Table Cont., 95\% of Table, 4},
        legend style={at={(0.5,-.7)},
        anchor=north,legend columns=4},
        xtick=data,
        x tick label style={text width=1cm},
        %%ylabel=Time (seconds),
        enlargelimits=0.05,
        ]
    \addplot
        coordinates {(5\% of Table Cont., 7.973) (5\% of Table, 7.973) (95\% of Table Cont., 9.123) (95\% of Table, 9.123) (4,1)}; %hash
    \addplot
        coordinates {(5\% of Table Cont., .7228) (5\% of Table, .7228) (95\% of Table Cont., 5.732) (95\% of Table, 5.732) (4,1)};	%small
        \addplot
        coordinates {(5\% of Table Cont., 0) (5\% of Table, 0) (95\% of Table Cont., 2.0047) (95\% of Table, 2.0047) (4,1)};	%large
    \addplot
        coordinates {(5\% of Table Cont., 1.413) (5\% of Table, 0) (95\% of Table Cont., 1.6313) (95\% of Table, 0) (4,1)};

    \legend{Hash, Small, Large, Cont.}
           \draw (37,108) node {\textasteriskcentered};
           \draw (106,880) node {* Our Choice};
           \draw (137,108) node {\textasteriskcentered};
           \draw (287,204) node {\textasteriskcentered};
           \draw (362.5,238) node {\textasteriskcentered};
           %\draw (63,40) node {\tiny{N/A}};
           %\draw (161,40) node {\tiny{N/A}};
           %\draw (187,40) node {\tiny{N/A}};
           %\draw (387,40) node {\tiny{N/A}};

    \end{axis}
    \end{tikzpicture}
    \caption{\small Our query planner picks the best algorithm for \texttt{SELECT} queries based on a first scan that determines which oblivious operator to use. Bars omitted when an algorithm is not applicable.}
    \label{figC3}
\end{figure}

\begin{figure}
\scriptsize
\begin{tabular}{l|rrr|rrr}
\multicolumn{7}{c}{\textbf{\rev{Join Performance -- 500 Rows Obliv. Mem.}}}\\
Table 1&\multicolumn{3}{c}{5,000 rows}&\multicolumn{3}{c}{10,000 rows}\\\hline
Table 2&\multicolumn{1}{c}{Hash}&\multicolumn{1}{c}{Opaque}&\multicolumn{1}{c|}{0-OM}&\multicolumn{1}{c}{Hash}&\multicolumn{1}{c}{Opaque}&\multicolumn{1}{c}{0-OM}\\
100&\textcolor{blue}{0.023s}&0.205s&\textcolor{red}{0.404s}&\textcolor{blue}{0.047s}&0.535s&\textcolor{red}{1.017s}\\
1,000&\textcolor{blue}{0.141s}&0.259s&\textcolor{red}{0.531s}&0.274s&\textcolor{blue}{0.553s}&\textcolor{red}{1.092s}\\
5,000&0.667s&\textcolor{blue}{0.529s}&\textcolor{red}{1.019s}&1.289s&\textcolor{blue}{0.822s}&\textcolor{red}{1.585s}\\
10,000&1.267s&\textcolor{blue}{0.808s}&\textcolor{red}{1.581s}&\textcolor{red}{2.592s}&\textcolor{blue}{1.340s}&2.497s\\
25,000&3.300s&\textcolor{blue}{2.078s}&\textcolor{red}{3.825s}&\textcolor{red}{6.540s}&\textcolor{blue}{3.046s}&5.337s\\
\end{tabular}\vspace{.5em}
\begin{tabular}{l|rrr|rrr}
\multicolumn{7}{c}{\textbf{\rev{Join Performance -- 7,500 Rows Obliv. Mem.}}}\\
Table 1&\multicolumn{3}{c}{5,000 rows}&\multicolumn{3}{c}{10,000 rows}\\\hline
Table 2&\multicolumn{1}{c}{Hash}&\multicolumn{1}{c}{Opaque}&\multicolumn{1}{c|}{0-OM}&\multicolumn{1}{c}{Hash}&\multicolumn{1}{c}{Opaque}&\multicolumn{1}{c}{0-OM}\\
100&\textcolor{blue}{0.007s}&0.044s&\textcolor{red}{0.202s}&\textcolor{blue}{0.015s}&0.154s&\textcolor{red}{0.533s}\\
1,000&\textcolor{blue}{0.016s}&0.053s&\textcolor{red}{0.244s}&\textcolor{blue}{0.031s}&0.165s&\textcolor{red}{0.571s}\\
5,000&\textcolor{blue}{0.050s}&0.149s&\textcolor{red}{0.520s}&\textcolor{blue}{0.103s}&0.335s&\textcolor{red}{0.792s}\\
10,000&\textcolor{blue}{0.095s}&0.334s&\textcolor{red}{0.794s}&\textcolor{blue}{0.192s}&0.431s&\textcolor{red}{1.282s}\\
25,000&\textcolor{blue}{0.241s}&0.938s&\textcolor{red}{2.041s}&\textcolor{blue}{0.479s}&1.040s&\textcolor{red}{2.869s}\\
%7500 Rows OM, 10,000\&25,000 row tables&.50s&1.10s&2.97s\\
%500 Rows OM, 1,000\&10,000 row tables&.28s&.54s&1.06s\\
%500 Rows OM, 10,000\&5,000 row tables&3.23s&2.08ss&3.80s\\
%500 Rows OM, 10,000\&25,000 row tables&6.34s&2.93s&5.15s\\
\end{tabular}
\caption{\small\rev{Foreign key joins with tables and oblivious memory of varying sizes. The fastest and slowest algorithm in each configuration are shown in blue and red, respectively. Reported number is average of 5 runs, standard deviation is always less than $8\%$ of average. Our planner picks the fastest algorithm for every entry in the table. }}
\label{joinfig}
\end{figure}

\subsection{Microbenchmarks}\label{optimizerEval}

\noindent\textbf{Impact of storage method}. 
Figure~\ref{figC1} compares our storage methods on various queries. Flat scans perform better when more rows are returned, but smaller queries perform much better with an index. Indexed \texttt{DELETE} and \texttt{UPDATE} queries outperform flat ones, but the fast flat \texttt{INSERT} query outperforms the indexed \texttt{INSERT}. The flat storage method's performance (outside of constant-time insertions) degrades linearly in table size, but point operations on indexes take polylogarithmic time. Figure~\ref{figScaling} shows how point queries scale. %For comparison, MySQL's point select latency was about .3ms for the 107,000 rows in the CFPB table. As mentioned above, the difference between MySQL and \name/ is not so significant compared to typical network latencies.

Often a combined table representation that maintains both storage methods for the same data proves effective. Although \name/ pays insertion and deletion costs for both methods, it can use the better representation for each query, an important benefit because many real-world workloads rely heavily on different kinds of reads. Figure~\ref{figworkload} shows \name/ running various workloads with flat, indexed, or both kinds of tables. One storage method alone sometimes performs well, but a combined representation often performs best.

%\subsection{Impact of Optimizer}\label{optimizerEval}
\noindent\textbf{Impact of query planner}. Figure~\ref{figC3} shows \name/'s choice of \texttt{SELECT} algorithms on queries that retrieve 5\% and 95\% of a 100,000 row table. The ``Hash'' algorithm is best asymptotically, but we pick an algorithm that performs $4.6$-$11\times$ better in practice.

\noindent\rev{\textbf{Join algorithm comparison}. Figure~\ref{joinfig} compares the performance of \name/'s join algorithms on foreign key joins for varying oblivious memory and table sizes. As explained in Section~\ref{optimizer}, input table sizes and oblivious memory are the only factors that affect join performance. Access to larger amounts of oblivious memory is particularly effective in speeding up the hash join algorithm because the size of the oblivious memory determines how many  times chunks of the first table need to be made into hash tables, which in turn determines the number of scans required of the second table. A large oblivious memory results in a join whose running time is almost linear in the size of the tables. For small oblivious memory, the performance behaves as expected of standard hash and sort-merge join algorithms: the hash join performs better for small tables but rapidly becomes worse than the sort-merge join as table sizes increase. }

\rev{The Opaque join always outperforms the variant that requires no oblivious memory because the two joins run effectively the same overall algorithm, with the Opaque join using oblivious memory to accelerate sorting. The 0-OM join gets faster as the amount of oblivious memory increases because of our optimization that does oblivious sorting inside the enclave when there is space available without compromising obliviousness (to save on enclave communication costs). As such, the algorithm gets faster with more enclave memory, regardless of whether the memory is oblivious.}

\noindent\textbf{Impact of padding mode}. 
Padding mode additionally hides the sizes of tables, intermediate results, and final outputs---comparable to the padding mode described but not evaluated by Opaque~\cite{ZDB+17}. We evaluate this mode by running queries on the CFPB table of 107,000 rows padded to 200,000 rows. Our aggregate query with the flat storage method had a 4.4$\times$ slowdown and a select had a 2.4$\times$ slowdown. The larger slowdown for aggregates results from the padding algorithm padding to the maximum supported number of groups for aggregates---in this case, 350,000. We did not evaluate padding mode for indexes as the benefit of indexes arises from knowledge of the selectivity of a query, the exact information padding hides. 
To our knowledge, no comparable system has implemented a pad mode, so we cannot compare to prior work. The results do, however, represent reasonable slowdowns for inflating a table's size by approximately 2$\times$ with padding. 

\section{Related Work}\label{related}
\noindent \textbf{Encrypted Databases}.
Fuller et al.~\cite{FVY+17} summarize prior work on cryptographically protected databases. The well-known CryptDB~\cite{PRZB12} enables a tradeoff between security and performance, encrypting fields differently according to security needs. Arx~\cite{PBP16} uses only strong encryption but leverages special data structures to allow search. Other solutions, including Demertzis et al., Sophos, and Diana \cite{DPP+16, Bost16, BMO17}, use searchable encryption.
Although all of these systems encrypt \emph{data}, they can leak \emph{access patterns}~\cite{IKK12, NKW15, CGPR15, ZKP16}. 

\noindent \textbf{SGX Databases}.
StealthDB~\cite{stealthDB} is a legacy-compatible, partially-oblivious database that does not provide integrity or hide access patterns to indexes. VeritasDB~\cite{veritasDB} provides integrity but not privacy. POSUP~\cite{POSUP} uses ORAM and SGX to search/update encrypted data and Cui et al.~\cite{CBZ+18} use SGX to speed up search over encrypted data, but both support a more limited range of functionalities than \name/. More recently, Oblix~\cite{oblix} builds an oblivious index that requires no obliviousness assumptions inside the enclave, and Obladi~\cite{obladi} considers concurrent ACID transactions but does not support indexes and only processes operations in batches over discrete time epochs. Opaque~\cite{ZDB+17} and Cipherbase~\cite{cipherbase} support only analytics queries that scan all the data, relying on oblivious sorts of an entire input table.
%To our knowledge, \name/ is the first oblivious database engine to efficiently support both transactional and analytics workloads.
%In particular, our variants of selection, aggregation and join operators not based on sorting (Section~\ref{oblivOps}) are novel over these prior oblivious systems, as is our combined use of oblivious B+ trees and flat storage.

EnclaveDB~\cite{enclavedb} is an SGX-based DBMS that does not hide access patterns. TrustedDB~\cite{trusteddb} uses older trusted hardware designs to build a protected database, but also does not protect access patterns. Many works also implement variations of other analytics systems on SGX~\cite{FBB+17,FVBG16,NFR+17,BEM+17}. M2R~\cite{DSC+15} and VC3~\cite{SCF+15} provide MapReduce and cloud data analytics functionalities, and HardIDX and LPAD~\cite{FBB+17, LPAD} build key-value stores that are not oblivious. 

\noindent \textbf{General-Purpose Oblivious Computing}.
ZeroTrace~\cite{SGF17} builds ORAM-based oblivious memory primitives over SGX, Pyramid ORAM~\cite{pyramid} builds an efficient ORAM for use in enclaves, and ObliVM \cite{oblivm} compiles oblivious versions of programs. By specializing data structures and operators for ORAM, \name/ outperforms na\"ive ORAM translations of database algorithms. Wang et al.~\cite{WNL+14} optimize data structures over ORAM, focusing on the case of recursive ORAM. 
Some of their techniques could complement our indexes when using a recursive ORAM position map. 
Roche et al.~\cite{RAC16} build a history-independent ``HIRB tree'' over an ORAM with variable-sized blocks, but do not support range queries. As seen in Section~\ref{HIRBcomp}, our indexes are up to 7$\times$ more efficient.

\rev{We use the Path ORAM~\cite{SDS+13} in our implementation, but any other ORAM could replace it with no other changes to the system. For indexed storage, where ORAM accesses dominate the cost of each operator, using a newer scheme such as Ring ORAM~\cite{ringoram} would result in performance improvements corresponding to the approximately $1.5\times$ improvement of Ring ORAM over Path ORAM. Unlike Ring ORAM, other ORAM optimizations designed for systems that provide cloud storage, such as Oblivistore~\cite{oblivistore}, CURIOUS~\cite{curious}, and TAOstore~\cite{taostore}, focus on reducing communication costs for the remote storage use case, which is less applicable in \name/, where the trusted and untrusted memory reside on the same device. }

\section{Conclusion}\label{conclusion}
\name/ closes the gap between previous enclave-based query processing engines and oblivious indexes by combining new oblivious query processing algorithms with accompanying data structures and an oblivious query planner. 
While obliviousness has a cost, \name/ approaches practical performance: it is competitive to 19$\times$ faster than Opaque~\cite{ZDB+17} and comes within 2.6$\times$ of Spark SQL. It also outperforms HIRB, a previous oblivious index structure, by over 7$\times$, completing point queries on a 1 million row table with 3.6--9.4ms latency. Our open source implementation of \name/ is available at {\small\url{https://github.com/SabaEskandarian/ObliDB}}.

%don't need this for submission
\section*{Acknowledgments}
We would like to thank Ankur Dave and Wenting Zheng for their assistance in reproducing the Opaque benchmarks, as well as Henry Corrigan-Gibbs for many helpful conversations. We would also like to thank the anonymous reviewers for several helpful comments and suggestions. 

This research was supported in part by affiliate members and other supporters of the Stanford DAWN project---Ant Financial, Facebook, Google, Infosys, Intel, Microsoft, NEC, SAP, Teradata, and VMware---as well as the NSF under CAREER grant CNS-1651570. Any opinions, findings, and conclusions or recommendations expressed in this material are those of the authors and do not necessarily reflect the views of the National Science Foundation.
Saba is supported by NSF, the DARPA/ARL SAFEWARE project, the Simons foundation, and a grant from ONR. The views expressed are those of the author and do not reflect the official policy or position of the Department of Defense, the National Science Foundation, or the U.S. Government.

\bibliographystyle{abbrv}
\bibliography{oramsgx}  % vldb_sample.bib is the name of the Bibliography in this case

%APPENDIX is optional.
% ****************** APPENDIX **************************************
% Example of an appendix; typically would start on a new page
%pagebreak

\begin{appendix}

\section{Security Theorem}\label{formalAppendix}
We model privacy by showing there exists a \emph{simulator} such that for all efficient adversaries~$\mathcal{A}$, $\mathcal{A}$ cannot distinguish between a real memory trace from \name/ and a memory trace from the simulator that is given access to query plans and table sizes. Since the simulator only sees what we intend to leak, the adversary cannot have learned any additional information from interacting with \name/. In this model, an (informal) theorem similar to that of Opaque~\cite{ZDB+17} also applies to \name/. Let $\mathcal{D}$ be a dataset, $\mathcal{S}$ be its schema, and $\mathcal{Q}$ be a query. Moreover, let \textsf{OPT($\mathcal{D}, \mathcal{Q}$)} be the choice of algorithms made by ObliDB's query planner for query $\mathcal{Q}$ on data $\mathcal{D}$ and \textsf{TRACE($\mathcal{D}, \mathcal{Q}$)} be the distribution of transcripts of memory accesses outside of oblivious memory made by ObliDB while running query $\mathcal{Q}$ on $\mathcal{D}$. Finally, $|\mathcal{D}|$ denotes the size of $\mathcal{D}$ and $|$\textsf{TRACE($\mathcal{D}, \mathcal{Q}$)}$|$ denotes the sizes of the memory traces of running each operator in $\mathcal{Q}$ on $\mathcal{D}$. Since ObliDB stores intermediate tables encrypted outside of the enclave, this includes intermediate table sizes.

\begin{theorem}
For all $\mathcal{D, S, Q}$, and security parameter $\lambda$, there is a poly-time simulator $\textsf{SIM}$ such that for all efficient adversaries~$\mathcal{A}$,
\begin{align*}
|\textsf{Pr[}\mathcal{A}&\textsf{(SIM}(|\mathcal{D}|, \mathcal{S}, \textsf{OPT}(\mathcal{D}, \mathcal{Q}), |\textsf{TRACE}(\mathcal{D}, \mathcal{Q})|))=1\textsf{]} \\&- \textsf{Pr[}\mathcal{A}(\textsf{TRACE}(\mathcal{D}, \mathcal{Q}))=1\textsf{] }| \leq \textit{negl}(\lambda).
\end{align*}
\end{theorem}

The fact that \textsf{SIM} exists means anything that can be learned by looking at the transcript of ObliDB running can also be learned by looking only at the sizes of the data/queries as well as the table schemas and physical operators chosen by the query planner. The theorem for padding mode replaces the data and trace size with a public parameter indicating the size to which we pad all tables.

To argue that SIM exists, we first argue that each operator output by \textsf{OPT} satisfies our obliviousness property. Next, we argue that the query planner's operations are oblivious with its only leakage being that inherent in the final choice of physical operator. We provide these arguments in Sections~\ref{oblivOps} and~\ref{optimizer}. With this, we have all the pieces required to explicitly describe \textsf{SIM} that prints an access pattern transcript distributed indistinguishably from $\textsf{TRACE}(\mathcal{D}, \mathcal{Q})$ because the trace of query $\mathcal{Q}$ on dataset $\mathcal{D}$ consists exactly of the accesses made by running the planner and then the chosen operator(s). 

\textsf{SIM} begins by reading $\mathcal{S}$ and $|\textsf{TRACE}(\mathcal{D}, \mathcal{Q})|$. It uses this information to simulate the access pattern of one scan over~$\mathcal{D}$. This is identical to the access pattern of the query planner. Now \textsf{SIM} reads \textsf{OPT}($\mathcal{D}, \mathcal{Q}$) to determine which operator to simulate. Using the provided choice of operator, the schema $\mathcal{S}$, and its knowledge of input and output table sizes gleaned from $|\textsf{TRACE}(\mathcal{D}, \mathcal{Q})|$, it simulates the access pattern described in the body of the paper for the selected operator on $\mathcal{D}$ (i.e. some number of linear scans or ORAM operations). This completes the simulated output which is distributed indistinguishably from that of \textsf{TRACE}($\mathcal{D}, \mathcal{Q}$). The simulator $\textsf{SIM'}$ for padding mode behaves analogously to $\textsf{SIM}$. 

\section{ORAM Formal Definition and Path ORAM}\label{oramAppendix}

Oblivious RAM (ORAM), a cryptographic primitive first proposed by Goldreich and Ostrovsky~\cite{GO96}, hides access patterns to data in untrusted storage. 
For our purposes, an ORAM consists of a small trusted \emph{client} which resides inside an enclave and performs reads/writes to untrusted memory accessible by the OS. 
Merely encrypting data still reveals access patterns to the data being requested or written, which can leak private information about the data~\cite{IKK12}. 
ORAM shuffles the locations of blocks in memory so repeated accesses to the same block and other patterns are hidden from the adversary. 
Specifically, ORAM guarantees that any two access patterns \emph{of the same length} are {computationally indistinguishable}. 

\ignore{ %remove for space saving if needed
\begin{definition}[ORAM Security\cite{SDS+13}]
Let $\overrightarrow{y}:=\textbf{((op$_1$,a$_1$,}$ $\textbf{data$_1$),}\textbf{...,(op$_M$, a$_M$, data$_M$))}$ denote a data request sequence of length $M$, where each \textbf{op$_i$} denotes a \textbf{read(a$_i$)} or a \textbf{write(a$_i$, data)} operation. Specifically, \textbf{a$_i$} denotes the address of the block being read or written, and \textbf{data$_i$} denotes the data being written.

Let $A(\overrightarrow{y})$ denote the (possibly randomized) sequence of accesses to the untrusted storage given the sequence of data requests $\overrightarrow{y}$. An ORAM construction is said to be secure if:
\begin{denseenumerate}
\setlength\itemsep{0pt}
\item For any two data request sequences $\overrightarrow{y}$ and $\overrightarrow{z}$ of the same length, their access patterns $A(\overrightarrow{y})$ and $A(\overrightarrow{z})$ are computationally indistinguishable by anyone but the client ORAM controller.

\item The ORAM construction is correct, in the sense that it returns, on input $\overrightarrow{y}$, output data that is consistent with $\overrightarrow{y}$ with probability $\geq 1 - \textit{negl}(|\overrightarrow{y}|)$. That is, the ORAM may fail only with a negligible probability $\textit{negl}(|\overrightarrow{y}|)$.
\end{denseenumerate}
\end{definition}
}

\noindent\textbf{Implementing ORAM}. \name/ uses the Path ORAM~\cite{SDS+13}, which operates by storing encrypted blocks of memory in a tree structure. Every read or write to a block (reads and writes are indistinguishable) reads a path from the root to a leaf, and then writes the same path again, regardless of where in the path the desired block sits. The contents of every node in the path are decrypted, read, and re-encrypted. To prevent leaking statistical information about repeated accesses to the same address, a block is randomly reassigned to a new part of the tree after each access. This causes ORAM reads and writes to incur an $O(\log N)$ overhead, where $N$ is the ORAM's size in blocks. If the tree lacks space to store some node in its designated place, the node is kept in an off-tree \emph{stash} until it can find space in a future operation. Path ORAM guarantees that the stash stays quite small with overwhelming probability. 

\noindent\textbf{Recursive vs Nonrecursive ORAM}. One feature of Path ORAM requires further discussion. In order to know which path down the tree to read to find a given block, the ORAM client keeps a \emph{position map} that maps each block of memory to a leaf in the tree that identifies the path where it can be found. Since the size of the position map is a fixed fraction of the size of the raw data, Path ORAM recursively stores the position map in a second ORAM and repeats until the client storage requirement becomes sufficiently small. We call an ORAM with no recursion a \emph{nonrecursive} ORAM and an ORAM that recursively uses a second ORAM a \emph{recursive} ORAM. In practice, because the size of an entry in a position map is many times smaller than a block of data, at most one layer of recursion suffices to store large quantities of data. For example, a 10MB position map in our implementation can support 1.1~million records (regardless of record size), and a 20MB position map can store twice as many records. Adding a second layer of recursion, where each of those 1.1 million records represent another 1.1~million records, comes at an approximately 2$\times$ performance overhead but allows the same 10MB position map to support 1.2~\emph{trillion} records. 

\noindent\rev{\textbf{Segmenting ORAM}. In addition to optimizing ORAM to minimize storage costs, we can also optimize to reduce computational costs. One way to do this is to separate one ORAM into multiple smaller ORAMs in a way that the choice of which ORAM is written to by a given operation does not leak any additional information. Although ORAM's computation costs scale logarithmically in the size of a given ORAM, dramatically reducing the size of an ORAM can still have a significant impact on performance. For example, \name/ uses a separate ORAM for each table because it does not hide which tables a query reads or modifies. This optimization could be taken further by using a separate ORAM for an index structure and the data for each table, or even using a separate ORAM for each level of a B+ tree (where padding would happen on a per-level basis rather than for the whole tree). These optimizations do not compromise obliviousness because the access patterns between levels of a B+ tree in a read, insert, or delete operation, once padded to the worst case scenario, are publicly known. The ORAM only needs to hide \emph{which entry} in a given level is accessed to preserve obliviousness.}

\end{appendix}

% ensure same length columns on last page (might need two sub-sequent latex runs)
\balance

\end{document}